\documentclass[oupdraft]{bio_arxiv}
\usepackage{url}
\usepackage{bm}
\usepackage{bbm}
\usepackage{xcolor}
\usepackage{ulem}
\usepackage[colorinlistoftodos]{todonotes}

\begin{document}

\title{
A copula-based set-variant association test for bivariate continuous or mixed phenotypes
}

\author{JULIEN ST-PIERRE$^{\ast}$\\[4pt]
\textit{Department of Epidemiology, Biostatistics and Occupational Health, McGill University, Montréal, Québec, Canada}
\\[2pt]
{julien.st-pierre@mail.mcgill.ca}\\[4pt]
KARIM OUALKACHA\\[4pt]
\textit{Département de Mathématiques, Université du Québec à Montréal, Montréal, Québec, Canada}}

\markboth%
{J. St-Pierre and K. Oualkacha}
{Multivariate rare-variant association for bivariate mixed outcome}

\maketitle

\footnotetext{To whom correspondence should be addressed.}

\begin{abstract}
{In genome wide association studies (GWAS), researchers  are  often  dealing  with non-normally distributed traits or a mixture of discrete-continuous traits. However, most of the current region-based methods rely on multivariate linear mixed models (mvLMMs) and assume a multivariate normal distribution for the phenotypes of interest. Hence, these methods are not applicable to disease or non-normally  distributed  traits. Therefore, there is a need to develop unified and flexible methods to study association between a set of (possibly rare) genetic variants and non-normal multivariate phenotypes. Copulas  are  multivariate  distribution  functions  with  uniform  margins on the $[0, 1]$ interval and they  provide  suitable  models  to  deal  with  non-normality of errors in multivariate association studies. We propose a novel unified and flexible Copula-Based Multivariate Association Test (CBMAT) for discovering association between a genetic region and a bivariate continuous or mixed phenotype. We also derive a data-driven analytic p-value procedure of the proposed region-based score-type test. Through simulation studies, we demonstrate that CBMAT has well controlled type I error rates and higher power to detect associations compared with other existing methods, for discrete and non-normally distributed traits. At last, we apply CBMAT to detect the association between two genes located on chromosome 11 and several lipid levels measured on 1,477 subjects from the ASLPAC study.}
{copulas; multivariate phenotype; mixed binary-continuous phenotypes; gene-based tests; rare-variant association; variance component score test.}
\end{abstract}

\section{Introduction}
\label{sec1}
Among the thousands of genetic variants that have been identified by genome wide association studies (GWAS), $4.6\%$ of single-nucleotide polymorphisms (SNPs) and $16.9\%$ of genes have proved to be associated with multiple correlated phenotypes (\citealp{solovieff2013}). This has led to a growing interest in discovering novel biology of pleiotropy by jointly analyzing multiple traits simultaneously. Pleiotropy occurs when several phenotypes are co-regulated by a common gene (\citealp{stearnsOneHundredYears2010,williamsPleiotropyNaturalSelection1957}). Some phenotypes are intrinsically multivariate, such as blood pressure, which is measured by both systolic and diastolic pressure. Moreover, analysis of multiple neurobiological and/or psychological genetically influenced quantifiable traits, i.e. endophenotypes (\citealp{gottesmanEndophenotypeConceptPsychiatry2003}), has been proposed to facilitate the search of susceptibility genes related to heterogeneous/complex traits, such as schizophrenia (SZ) and major depressive disorder (MDD) (\citealp{hall2010,iacono2017}). Indeed, multiple endophenotypes often measure the same underlying trait and might bear a more direct relationship with the biological etiology of a clinical disorder. Ignoring the fact that multiple phenotypes are related and might share a common genetic basis of complex traits, the single-trait study paradigm is likely to suffer from a potential loss of power and may provide less accurate prediction of some phenotypes compared to simultaneous analysis of multiple traits (\citealp{yang2016}). 

Region-based association methods have become increasingly popular given large-scale genomic data provided by next generation sequencing technology and are often of interest in genome wide association studies (GWAS). Such methods aggregate variants into groups based on pre-defined criteria, such as gene boundaries or linkage-disequilibrium (LD) blocks, and then test for phenotypic effects within each group. Region-based methods reduce the multiple testing burden of single-variant analyses. Furthermore, they can increase power by combining information across biologically meaningful units (\citealp{petersenAssessingMethodsAssigning2013}). Several approaches have been developed to handle multivariate association in both single-variant and region-based frameworks. For association methods between a single genetic variant and a multivariate phenotype, a nice overview can be found in \citet{yang2012} (see also \citealp{schaidStatisticalMethodsTesting2016,shrinerMovingSystemGenetics2012,solovieff2013, zhangTestingAssociationMultiple2014}). In the region-based framework, many strategies have been explored for testing multivariate association and can be classified into three main categories: regression analysis, variable reduction analysis, and combining approach. 

The combining approach consists in aggregating test statistics from either multiple single-variant-based tests (\citealp{vandersluisMGASPowerfulTool2015}) or test statistics obtained from single-variant-based multivariate tests (\citealp{guo2013}) to obtain one overall, multivariate gene-based p-value. This approach can lead to low power because it discards the additional information obtained by combining many outcomes into a unified analysis. Indeed, modeling multivariate phenotypes may increase the power over analyzing individual phenotypes separately (\citealp{klei2008}). The second category includes variable reduction methods such as principal component analysis (PCA) and canonical correlation analysis (CCA). CCA, which extracts the linear combination of traits that explain the largest possible amount of covariance between two sets of variables, was initially proposed for multivariate association analysis by \citet{ferreiraMultivariateTestAssociation2009} and later extended to perform multivariate gene-based association tests (\citealp{tangGenebasedTestAssociation2012, seoaneCanonicalCorrelationAnalysis2014}). Important limitations of CCA is that these methods cannot accomodate covariates and are not appropriate for the analysis of non-normally distributed traits. To this date, work is still needed to extend the PCA framework for joint analysis of multiple phenotypes in GWAS to multiple phenotype analysis in region-based association studies (\citealp{liu2019}).

Finally, the last category regroups methods based on regression analysis. The use of functional linear models for the analysis of multiple variants and multiple traits has been proposed by \citet{wangPleiotropyAnalysisQuantitative2015}. However, most of the regression methods are based on multivariate linear mixed models (mvLMMs) which assume multivariate normal distribution for multiple phenotypes and a random effect to account for the cross-phenotype correlations (\citealp{MURAT,Dutta2019,MAAUSS}). A limitation of all these methods is that they are not applicable to disease or non-normally distributed traits. \citet{MAAUSS} extended their Multivariate Association Analysis using Score Statistics (MAAUSS) method so that it could also handle correlated binary and continuous phenotypes using the generalized estimating equation (GEE) framework. In practice, researchers are often dealing with both non-normally distributed traits and a mixture of discrete-continuous traits, hence there is a need to develop unified and flexible methods to study association between a set of rare/common variants and non-normal multivariate phenotypes.

 Copulas are multivariate distribution functions with uniform margins on the $[0,1]$ interval (\citealp{Nelsen}). They provide flexible models for joint distributions by separating specification of the marginal distributions and the dependence. In other words, they allow for modelling a dependency structure between several random variables regardless of their marginal distributions. Hence, they provide suitable models to deal with non normality of errors in multivariate association studies. Copulas are extensively used in regression models due to their great flexibility, as they can be incorporated in models with binary outcomes (\citealp{Genest}) as well as with mixed discrete and continuous outcomes (\citealp{Leon}). More recently, they have been incorporated into a single-marker association test to increase the power of genetic association studies of rare variants (\citealp{doi:10.1002/gepi.22265}) and into a gene-based association study for bivariate survival traits based on functional regression (\citealp{Wei2019}). Of note, the latter model is only valid for continous survival traits and the authors have assumed the same effects for the covariates and the genetic region on both marginal survival times, which is unrealistic in practice when dealing with different diseases for instance.

In this paper, we propose a novel unified and flexible copula-based multivariate test, CBMAT (Copula Based Multivariate Association Test), for discovering association between a genetic region and a bivariate continuous or mixed phenotype. By using copulas to model the joint distribution of phenotypes, CBMAT allows a wide range of continuous and discrete marginal distributions for the phenotypes. Many of the existing multivariate methods for region-based tests can be viewed as special cases of CBMAT with a Gaussian copula and particular choices of kernels to relate genetic variants to multiple phenotypes. Finally, we also derive a data-driven analytic p-value procedure of the proposed region-based score-type test.

The article is organized as follows: in Section \ref{sec2}, we introduce our proposed method CBMAT and derive a data-driven analytic p-value procedure of the proposed region-based score-type test. In Section \ref{sec3} and Section \ref{sec4} respectively, we describe our simulation studies and demonstrate that CBMAT has well controlled type I error rates and higher power to detect associations compared with other existing methods, for discrete and non-normally distributed traits. At last, we apply CBMAT to detect the association between two genes located on chromosome 11 and several lipid levels measured on 1,477 subjects from the ASLPAC study.

\section{Material and Methods}\label{sec2}
Our goal in this section is to develop a unified and flexible multivariate mixed-effects-model for a wide range of discrete and continuous distributions. First, we present the copula-based model for the joint density of $(Y_1, Y_2)$ when both traits are continuous and when one trait is discrete binary. Secondly, we briefly review some existing multivariate linear mixed models for region-based association. Third, we present our proposed CBMAT score statistics for testing the variance component of the region effect. Finally, we derive the null distribution of our proposed score statistics and its data-driven analytic p-value procedure.
\subsection{Copula-based model for joint density}
Assume we observe 2 traits for $N$ independent subjects, $\bm{Y}_i=(y_{i1},y_{i2})^T$, $i=1,...,N$, $m$ covariates, $\bm{X_i}=(x_{i1},x_{i2},...,x_{im})^T$, and a set of SNPs, $\bm{G_i}=(g_{i1},g_{i2},...,g_{ir})^T$ that contains $r$ biallelics variants coded $0,1,2$, representing the number of minor alleles. The proposed model assumes that the (conditional) marginal distributions of $Y_{ik}$, $k=1,2$, are belonging to the exponential family, and it links $\mu_{ik}$, the conditional mean of $Y_{ik}$, to $\bm{X}_i$ and $\bm{G}_i$ via a (conditional) marginal generalized model, as follows
\begin{equation}
\begin{aligned}
\label{marginalmean}
&\text{E}[Y_{ik}|\bm{X}_i,\bm{G}_i]=\mu_{ik}=g_k^{-1}\left(\bm{X}_i^T\bm{\gamma}_k+\bm{G}_i^T\bm{\beta}_k\right),\\
&\text{Var}[Y_{ik}|\bm{X}_i,\bm{G}_i]=\phi_k\cdot\nu_k(\mu_{ik}),
\end{aligned}
\end{equation}
where, for the $k^{th}$ trait, $\bm{\gamma_}k=(\gamma_{k1},...,\gamma_{km})^T$ and $\bm{\beta_}k=(\beta_{k1},...,\beta_{kr})^T$ are vectors of regression coefficients for $\bm{X_}i$ and $\bm{G_}i$ respectively, $g_k(.)$ is the link function, $\phi_k$ a dispersion parameter and $\nu_k$ the variance function. Thus, the traits (conditional) marginal cumulative distribution functions, $F_k(y_{ik}|\bm{X}_i, \bm{G}_i)$, are completely specified. We then specify the joint CDF of $Y_{i1}$ and $Y_{i2}$ as follows 
\begin{equation}
\label{joint-CDF}
\bm{F}\left(y_{i1},y_{i2} \mid \bm{X}_i, \bm{G}_i \right)=C_{\theta}\left\{F_{1}\left(y_{i1} \mid \bm{X}_i, \bm{G}_i \right), F_{2}\left(y_{i2} \mid \bm{X}_i, \bm{G}_i \right)\right\},
\end{equation}
where $C_{{\theta}}$ is a copula function, which accounts for dependence between marginal CDFs $F_1, F_2$ through the parameter ${\theta}$; i.e., $C_{{\theta}}$ is a joint distribution function in the unit square
\begin{eqnarray*}
C_{\bm{\theta}}: [0,1]^{2} &  \longmapsto &  [0,1] \\
             (u_1, u_2) &\longmapsto & C_{\bm{\theta}}(u_1,u_2).
\end{eqnarray*}
In the sequel, we drop the conditional notation with respect to $\bm{X}_i$ and $\bm{G}_i$ for sake of clarity.
For two continuous phenotypes, the (conditional) joint density function is given as 
\begin{align}
\label{continuous-traits}
f(y_{i1},y_{i2})=f_1(y_{i1})f_2(y_{i2})c_{\theta}(F_1(y_{i1}),F_2(y_{i2})),
\end{align}
where $c_{\theta}(u,v) = \partial^{2} C_{\theta}(u, v) / \partial u \partial v$ stands for the copula density. In the mixed phenotype case, that is for $Y_{i1}$ binary and $Y_{i2}$ continuous, a liability threshold model which transforms the binary trait into a probit latent marginal variable is proposed. We define $Y_{i1}^*$ such that it has a normal CDF $F_1^{*}(y_{i1}^*)=\Phi\left(\frac{y_{i1}^*-g_1(\mu_{i1})}{\sigma_1}\right)$ (i.e., $Y_{i1}^{*}\sim N(g_1(\mu_{i1}), \sigma_1^2)$), with $Y_{i1}$ and $Y_{i1}^{*}$ linked in a way such that 
$Y_{i1}=I(Y_{i1}^{*}> y^{c})$, where $I(\cdot)$ is the indicator function and $y^{c}$ is a cutpoint. For convenience and identifiability reasons, we take $y^{c} = 0$ and $\sigma_{1}^2=1$. Letting $\bm{F}^*$ be the joint CDF of $Y_{i1}^*$ and $Y_{i2}$, we define the following probability of event for $Y_{i1}$ and $Y_{i2}$ as follows
\begin{align*}
P(Y_{i1}=y_{1},Y_{i2}<y_{2})=
\begin{cases}
\bm{F}^*(0,y_2) & \text{if } y_1=0, \\
F_2(y_2)-\bm{F}^*(0,y_2) & \text{if } y_1=1.
\end{cases}
\end{align*}
Let now $\bm{F}^*$ be determined by a copula model, as in (\ref{joint-CDF}). This leads to $$\bm{F}^{*}( y_{i1}^{*}, y_{i2} )=C_\theta\big\{ F_{1}^{*}(y_{i1}^{*}), F_2(y_{i2})\}.$$ Thus, one can write the joint (conditional) density $f\big(y_{i1},y_{i2}\big)=\partial P(Y_{i1}=y_{i1}, Y_{i2}<y_{i2})/\partial y_2$ as 
\begin{equation}
\label{mixed-traits}
f\big(y_{i1},y_{i2}\big)=f_2(y_{i2})\times\Big[C_{\theta}^{01}\{ 1-\mu_{i1}, F_2(y_{i2}) \}\Big]^{I(y_{i1}=0)} \times\Big[1-C_{\theta}^{01}\{ 1-\mu_{i1}, F_2(y_{i2}) \} \Big]^{I(y_{i1}=1)}, 
\end{equation}
where $\mu_{i1}$ is given in (\ref{marginalmean}), $f_{2}$ is the density function of $Y_2$, and $C_{\theta}^{01}$ is the derivative of the copula function $C_\theta$ over its second argument, $C_{\theta}^{01}(u,v)=\partial
C_{\theta}(u,v)/\partial v$. 

Recall that the regression parameters linking both covariates and genotypes to the phenotypes are kept in the marginal CDFs, and so, they are marginally meaningful. Moreover, both between- and within-trait polygenic heritability (i.e. polygenic dependence, (\citealp{Bauman2005})) are “margin-free”, in a way that they are characterized by the copula alone; this is another advantage of the proposed model.

\subsection{Existing mvLMM approaches}
In the multivariate set-based association framework, the overall effect of the genomic region on the two phenotypes, $\bm\beta=(\bm\beta_1^{\top},\bm\beta_2^{\top})^{\top}$, is assumed to be a vector of (normally-distributed) random effects with mean zero and variance-covariance matrix $\eta \Sigma_P \otimes \Sigma_G$, where $\otimes$ is the Kronecker product, $\eta$ is an unknown variance component, $\Sigma_P$ is a $2 \times 2$ matrix that captures the relationship among the effect sizes of each variant on the $2$ phenotypes, and $\Sigma_G$ is a $r \times r$ matrix that models the relationship among the effect sizes of the variants on each phenotype.

In our proposed model, we also adopt the random effect assumption of the region overall-effect on the two phenotypes. Consequently, if one assumes marginal normal distributions for the two phenotypes and set $C_{{\theta}}$ to be a Gaussian copula, then the proposed model (\ref{joint-CDF}) will be equivalent to the multivariate mixed-effects-model of MURAT (\citealp{MURAT}) and Multi-SKAT (\citealp{Dutta2019}), which assume multivariate normality of the error terms, and thereby Gaussian distributions for the univariate margins. Thus, in comparison with existing methods, the proposed model allows for marginal CDFs to be any distribution belonging to the exponential family and can handle both continuous and discrete phenotypes simultaneously.

Of note, the main difference between most existing mvLMM approaches results from different specifications of the two kernel matrices, $\Sigma_G$ and $\Sigma_P$. For instance, MURAT assumes a common correlation, $\rho$, for the effects of the same variant on different phenotypes and also that the effects of different variants are uncorrelated; i.e. MURAT assumes that ${\Sigma_P=(1-\rho)I_K+\rho\bm{1_K}\bm{1_K}^T}$, where $\bm{1_K}$ is a $K$-dimensional vector of ones. The Multi-SKAT method (\citealp{Dutta2019}) suggests different choices for $\Sigma_P$. For instance, if we assume that the effect sizes of a variant $j$ are the same for all the phenotypes, in which case $\beta_{1j} = \ldots = \beta_{Kj}$, the homogeneous kernel is $\Sigma_P = \bm{1_K}\bm{1_K}^T$ (i.e. $\rho =1$ in $\Sigma_P$). On the other hand, if we assume that the effect sizes $(\beta_{1j},...,\beta_{Kj})$ are uncorrelated among themselves, the heterogeneous kernel is $\Sigma_P= I_K$ (i.e. $\rho =0$ in $\Sigma_P$). Other suggestions for $\Sigma_P$ are given in detail in \citet{Dutta2019}. 

The choice of $\Sigma_G$ has been extensively studied in the literature (\citealp{SKAT,leeOptimalUnifiedApproach2012}) and a common choice is to assume $\Sigma_G = \bm{W}$, where  $\bm{W}=\text{diag}(w_1,...,w_r)$ is an $r \times r $ diagonal matrix of the weights to be used for the r variants. Typically, the weights are chosen such that they are inversely proportional to the minor allele frequency (MAF) to up-weight rare variants, for instance by using $w_j = beta(MAF_j, 1,25)$.

\subsection{Hypothesis testing and parameter estimation}
\label{hypothesis testing}
Under the copula-based model (\ref{joint-CDF}), the association between a region that containts $r$ variants and the bivariate phenotype can be tested by evaluating the null hypothesis $H_0: \bm{\beta} = \bm{0}$. As mentionned in the previous section, we assume $\bm{\beta}$ is a random effects vector that follows an arbitrary probability distribution
\begin{align}
\bm\beta=\left(\bm\beta_1^T,\bm\beta_2^T\right)^T\sim \bm{H}\left(\bm{0}_{2r},\bm{\Sigma}_{\beta}=\eta \Sigma_P \otimes \Sigma_G\right).\label{eq. method.6}
\end{align} 
In this context, testing $\bm{\beta} = \bm{0}$ corresponds to testing $\eta = 0$. In what follows, we derive a score-type test statistic for the null hypothesis $H_0: \eta = 0$. 
First, one can see that the conditional likelihood function given $\bm{\beta}$ is given as
\begin{align}
\label{condit.lik}
L(\bm{\xi}|\bm{\beta})&=\prod_{i=1}^{n}f(y_{i1},y_{i2}),
\end{align}
where $f(y_{i1},y_{i2})$ is given by (\ref{continuous-traits}) when both traits are continuous, and by (\ref{mixed-traits}) for the mixed trait case, $\bm{\xi}=(\theta,\bm{\gamma_}1,\bm{\gamma_}2,\phi_1,\phi_2)^T$, where $\theta$ is the copula parameter, while  $\bm{\gamma_}1, \bm{\gamma_}2, \phi_1$, and $\phi_2$ are the parameters of the marginal GLMs given in (\ref{marginalmean}). Of note, the product factorization of the conditional likelihood in (\ref{condit.lik}) results from the fact that, conditioning on $\bm{\beta}$, the subjects are independent. The full likelihood is given by the $2r$-dimensional integral
\begin{align}
\label{eq. method.13}
L(\eta, \bm{\xi})=\int_{\bm{\beta}}L(\bm{\xi}|\bm{\beta}) \ \bm{H}(\bm{\beta}) d\bm{\beta}.
\end{align}
Finally, the score test for $H_0$: $\eta=0$ is based on the score statistic
\begin{align*}
U(\eta)=\frac{\partial}{\partial\eta}\text{log }L(\eta,\bm{\xi}).
\end{align*}
The problem with the direct computation of this score is the evaluation of the integral in dimension $\mathbbm{R}^{2r}$ in \eqref{eq. method.13}.
To solve this computational problem, we rely on the work proposed by (\citealp{Taylor}) and approximate
the integral using Taylor's expansion techniques of $L(\bm{\xi}|\bm{\beta}) $ in the neighborhood of $\bm{\beta}=\bm{0}_{2r}$. 
In Appendix A, we show that
\begin{align}
\label{eq:score}
U(\eta) \approx  \frac{1}{2} \Big( \bm{L}^T (\Sigma_P \otimes \bm{G} \Sigma_G \bm{G}^T) \bm{L} +\text{tr}\left\{ (\Sigma_P \otimes \bm{G}\Sigma_G \bm{G}^T) \bm D\right\} \Big), 
\end{align}
where $\bm{L} = (\bm{L}_1^T,\bm{L}_2^T)^T$ is a $2n \times 1$ vector, with $\bm{L}_1 = (\bm{L}_{11},\ldots,\bm{L}_{1n})^T$ and $\bm{L}_2 = (\bm{L}_{21},\ldots,\bm{L}_{2n})^T$, such that 
$$
\frac{\partial }{\partial\bm{\beta}}l(\bm{\xi}|\bm{\beta}) = \sum_{i=1}^n \begin{bmatrix}\bm{L}_{1i} \\[1em] \bm{L}_{2i}\end{bmatrix}  \otimes \bm{G}_i = \left(I_2\otimes \bm{G}^T\right) \bm{L},
$$
and $\bm G$ is the $n\times r$ matrix of genotypes, $\bm D$ is a block matrix defined in Appendix A, whose entries are corresponding to the Hessian of the (conditional) log-likelihood with respect to $g(\bm \mu)$, such that 
\begin{equation}
\label{eq:hessian}
\frac{\partial^2}{\partial\bm{\beta}\partial\bm{\beta}^T}l(\bm{\xi|\beta})=\left(I_2\otimes \bm{G}^T\right) \bm D \left(I_2\otimes \bm{G}\right).
\end{equation}
The score statistic in (\ref{eq:score}) depends on the data and the unknown vector parameter $\bm{\xi}$ through $\bm L$ and $\bm D$. It depends also on the two kernel matrices $\Sigma_P$ and $\Sigma_G$, which need to be specified based on prior knowledge beforehand. Thus, our plug-in score test statistic is obtained by substituting the unknown entries of $\bm{\xi}$ in (\ref{eq:score}) by their maximum likelihood estimators under the null hypothesis $H_0: \eta=0$.

\subsection{Null distribution correction of the proposed plug-in score statistic} Under the null model, when the unknown parameters are fixed at their true values, standard asymptotic theory gives that $\left(I_2\otimes \bm{G}^T\right) \bm{L}$ follows an $2r$-variate normal distribution with mean zero and variance-covariance matrix $-E\left(\left(I_2\otimes \bm{G}^T\right) \bm D \left(I_2\otimes \bm{G}\right)\right)$. Thus, the asymptotic distribution of $U(\eta)$ is equivalent to that of $\sum_{i=1}^{2r}\lambda_i(\chi^2_1 - 1)/2$, where $\lambda_i$, $i=1,...,2r$, are eigenvalues of $\bm{B}^{1/2}(\Sigma_P \otimes \Sigma_G)\bm{B}^{1/2}$, with $\bm B = -\left(I_2\otimes \bm{G}^T\right) \bm D \left(I_2\otimes \bm{G}\right)$ is the negative of the Hessian matrix defined in \eqref{eq:hessian}. However, ignoring variability due to the fact that we are using estimates of $\bm{\xi}$, in the asymptotic variance-covariance matrix, $\bm B$, can lead to severe type I error inflation of the proposed score test. In our work, we rely on (\citealp{inverse}) and suggest a corrected variance-covariance matrix, say $\tilde{\bm B}$, of the score vector, which accounts for the variability induced by the plug-in estimates. This corrected matrix satisfies $\tilde{\bm B}^{-1} = \bm B^{-1} + \bm C_{\hat{\xi}}$, where $\bm C_{\hat{\xi}}$ is a $2r \times 2r$ positive definite symmetric matrix, which depends on $\hat{\bm \xi}$ and the data. Consequently, the corrected null distribution of the score statistic becomes $\sum_{i=1}^{2r}\tilde{\lambda}_i\chi^2_1/2-\sum_{i=1}^{2r}\lambda_i/2$, where $\tilde{\lambda}_i$, $i=1,...,2r$, are eigenvalues of $\tilde{\bm{B}}^{1/2}(\Sigma_P \otimes \Sigma_G)\tilde{\bm{B}}^{1/2}$. Finally, one can approximate the score test distribution under the null using the Davies method (\citet{davies1980}), and obtain the corresponding p-value. Appendix B gives more details about the derivation of the asymptotic distribution of the proposed score statistic, and outlines steps to obtain the corrected asymptotic variance-covariance matrix $\tilde{\bm B}$. 

In this work, we set $\Sigma_G = \bm{W}$, an $r\times r$ diagonal  matrix  of weights. Several choices of $\Sigma_P$ and their effect from the modelling perspective have been investigated in \citet{Dutta2019}. If a selected $\Sigma_P$ does not reflect biology, the test may have substantially reduced power. To avoid such a drawback and achieve robust power, we follow the same procedure as in MURAT, and set $\Sigma_{\rho} := \Sigma_P = (1-\rho)I_K+\rho\bm{1_K}\bm{1_K}^T$, then we aggregate results across different $\Sigma_{\rho}$ by selecting, from a grid of values over $[0,1]$, the optimal value of $\rho$ which maximizes the power of the test. 

\subsection{Data-driven analytic p-value procedure} We outline here our procedure for obtaining an analytic p-value of the proposed data-driven score test.
To do so, let $(\rho_1,\rho_2,\ldots,\rho_b)$ be a grid of values of $\rho$ in $[0,1]$, and let $U_{\rho_j}(\eta)$ the corresponding plug-in estimate score statistic. Deriving p-value for the minimum p-value statistic, based on $U_{\rho_j}(\eta)$'s, can be achieved using permutation or perturbation to calculate the Monte‐Carlo p-value, which is computationally expensive. Resampling-based techniques, which require less resampling steps, as in Multi-SKAT, can also be used to approximate the joint null distribution of $\left(U_{\rho_1}(\eta), \ldots, U_{\rho_b}(\eta)\right)$, and then one can approximate p-value for the minimum p-value statistic. Here, again, we rely on copulas to develop an analytic p-value procedure, which does not require permutation nor resampling techniques. In fact, noticing that the second term in the right-hand side of (\ref{eq:score}) has small variability compared to the first term, we consider a series of statistics 
\begin{align}
\nonumber
Q_{\rho_j} &= \bm{L}^T (\Sigma_{\rho_j} \otimes \bm{G} \Sigma_G \bm{G}^T) \bm{L} \\
           & = \bm{Z}^T \tilde{\bm{K}}_{\rho_j} \bm{Z},
\label{eq:scoreQ}
\end{align}
where $\bm{Z} = \tilde{\bm{B}}^{-1/2} \left(I_2\otimes \bm{G}^T\right) \bm{L} \sim N(\bm 0, \bm I_{2r})$, and $\tilde{\bm{K}}_{\rho_j} = \tilde{\bm{B}}^{1/2}(\Sigma_{\rho_j} \otimes \Sigma_G)\tilde{\bm{B}}^{1/2}$ for $j=1,\ldots, b$. Again, one can verify that the null distribution of $Q_{\rho_j}$ is $\sum_{i=1}^{2r}\tilde{\lambda}_i^{(j)}\chi^2_1$, where $\tilde{\lambda}_1^{(j)}, \ldots, \tilde{\lambda}_{2r}^{(j)}$ are eigenvalues of $\tilde{\bm{K}}_{\rho_j}$. The p-value of such a test statistic is $p_j = P(Q_{\rho_j} > q_{\rho_j}) = S_j(q_{\rho_j})$, where $S_j(.)$ and $q_{\rho_j}$ are the survival and the observed value of $Q_{\rho_j}, j=1,\ldots,b$. Now, assume $\mathcal{P}_{min} = min(P_1, \ldots,P_b)$, with $P_{j} = S_j(Q_{\rho_j})$. Our data-driven procedure rejects the null hypothesis if $\mathcal{P}_{min}$ is large, and its corresponding p-value is $P(\mathcal{P}_{min} < p_{min})$, where $p_{min}$ is the observed value of $\mathcal{P}_{min}$, which is given as follows 
\begin{align}
\nonumber
P\left(\mathcal{P}_{\min }<p_{\min }\right)&=1-P\left\{P_{1}>p_{\min}, \ldots, P_{b}>p_{\min }\right\}\\
\nonumber
&=1-P\left\{P_{1}<1-p_{\min }, \ldots, P_{b}<1-p_{\min }\right\} \\
\label{min.pval}
&=1-P\left\{Q_{\rho_1}>S_{1}^{-1}\left(1-p_{\min }\right), \ldots, Q_{\rho_b}>S_{b}^{-1}\left(1-p_{\min }\right)\right\}.
\end{align}
In order to derive a closed-form expression for the p-value in (\ref{min.pval}), we approximate the unknown joint null distribution of $(Q_{\rho_1},\ldots,Q_{\rho_b})$ by copulas. More precisely, we assume 
$$
P\left(Q_{\rho_1}>q_{\rho_1}, \ldots, Q_{\rho_b}>q_{\rho_b}\right)=C_{\Gamma}\left\{S_{1}\left(q_{\rho_1}\right), \ldots, S_{b}\left(q_{\rho_b}\right)\right\},
$$
where $C_{\Gamma}$ is a Gaussian copula indexed by a correlation matrix $\Gamma$. Under such an assumption, we have the following closed-form expression for the p-value
\begin{align}
\nonumber
P\left(\mathcal{P}_{\min }<p_{\min }\right) = 1-C_{\Gamma}\left(1-p_{\min }, \ldots, 1-p_{\min }\right).
\end{align}
The p-value calculation procedure is completed by specifying the correlation matrix $\Gamma$. In this work, we take advantage of the fact that, for $1\leq j,l \leq b$, we have (\citealp{Magnus1978})
\begin{equation}
\label{cov.Qjl}
Cov(Q_{\rho_j}, Q_{\rho_l}) = 2\text{tr}(\tilde{\bm{K}}_{\rho_j}\tilde{\bm{K}}_{\rho_l}),
\end{equation}
and we set $\Gamma$ to be the Pearson correlation matrix of $(Q_{\rho_1}, \ldots, Q_{\rho_b})$.

Of note, Kendall's tau correlations might be more suitable as entries of $\Gamma$, however analytical derivation of such correlations from (\ref{cov.Qjl}) requires solving $b(b-1)/2$ double-integrals equations (\citealp{sunMultivariateAssociationTest2019}). In our simulation studies in Section \ref{sec3}, we show that the use of Pearson correlation matrix to model dependence between the test statistics has no impact on type I error. 
Moreover, in Appendix D of the Supplementary Materials, we compare CBMAT p-values obtained using these two strategies, that is, either an approximation of the matrix $\Gamma$ using the rank-based dependence between the test statistics (i.e. Kendall's tau coefficients) calculated based on a resampling procedure, or using the Pearson correlation matrix, as described above. We show that these two procedures yield similar power for both continuous and mixed bivariate phenotypes.

\section{Simulation Studies}
\label{sec3}
We carried out simulation studies to evaluate and compare the proposed methodology for region-based association testing of bivariate phenotypes. Comparisons were made with MURAT and Multi-SKAT, using available packages in \texttt{R}. For Multi-SKAT, we assumed that effect sizes of a variant on different phenotypes were either homogeneous (Multi-SKAT Hom) or uncorrelated (Multi-SKAT Het).

{\it Data generation:} Genotype data of subjects with European ancestry from the 1000 Genomes Project (\citealp{1000G}) were used to mimic real data situations. From this public database, we extracted genetic variants within 500kbs of the BRCA1 gene, for which several known mutations are associated with a higher risk of developing breast, ovarian and prostate cancers (\citealp{BRCA1}). To avoid multicollinearity issues, we pruned variants in strong LD ($r^2>0.7$). 
Simulated data of $N=503$ subjects were generated according to scenarios defined by varying combinations of values for parameters (Kendall's $\tau$, $\rho$, $\eta$), for a total of 21 scenarios: 3 × 1 × 1 = 3 scenarios under $H_0$, and 3 × 3 × 2 = 18 scenarios under the alternative hypothesis, $H_a$. For a given scenario and for each individual, $i$, independently, we simulated the data according to the following steps:\\
1. $\mathbf{G}_i$ is the genotype vector of a randomly selected genetic region of $r = 30$ consecutive common/rare variants from the extracted large region around {\it BRCA1} gene (the selected region is the same for all subjects); \\
2. $\mathbf{X}_i =(1, X_{i1},X_{i2})^T$, with $X_{i1} \sim \textit{Bernoulli}(0.5)$ and $X_{i2} \sim \textit{N}(0,1)$;\\
3. $\bm{\beta}=\begin{pmatrix}\bm{\beta_}1\\\bm{\beta_}2\end{pmatrix}\sim N_{2r}\left(\bm{0}_{2r},\eta\begin{bmatrix}\bm{W}&\rho\bm{W}\\\rho\bm{W}&\bm{W}\end{bmatrix}\right)$;\\
4. $(U_i, V_i) \sim C_{\theta}$\\
5. $Y_{i1}=F^{-1}_{1}(U_i|\mu_{i1},\phi_1, \tau)$, and  $Y_{i2}=F^{-1}_{2}(V_i|\mu_{i2},\phi_2, \tau),$\\
with $\mu_{ik}=g_k^{-1}(\bm{X}_i^T\bm{\gamma}_k+\bm{G}_i^T\bm{\beta}_k),$
where $F_k^{-1}(\cdot)$ and $g_k^{-1}(\cdot)$ are respectively the inverse of the CDF function and the inverse link function of $Y_k$, $k=1,2$.

We fixed the covariates' effects at $\bm{\gamma}_1=(-0.20,0.33,0.78)^T$ and $\bm{\gamma}_2=(1.52,1.25,1.86)^T$, and we set the dispersion parameters to $\phi_1=\phi_2=1$. We fixed the weights matrix $\bm{W} = \bm{I}_r$. The dependency structure between $Y_1$ and $Y_2$ was induced using a selected Kendall's $\tau$ value and either a Gaussian, a Frank, or a Clayton copula. Our selected $\tau$ values ($0.05, 0.2, 0.4$) correspond to weak, moderate, and strong phenotype dependency, respectively. Under $H_0$, we set $\eta = 0$. Under $H_a$, the value of $\eta$ depends on the fraction of causal variants $v$ and on the traits heritability $h^2$, which is defined as the fraction of the total phenotypic variability that is attributable to genetic variability. We set $v$ equals to $10\%$ and $20\%$ respectively, and $h^2$ equals to 2$\%$. Of note, as the binary trait was generated based on a latent probit model, heritability was calculated for its underlying latent continuous trait. As there exists no closed form relationship between $\eta$, $v$ and $h^2$ for non-linear models, calculations were made based on the parameter relationships derived for the normal traits. More details are given in Appendix C of the Supplementary Materials. Under $H_a$, we varied the region-specific pleiotropy parameter as $\rho=(0,0.4,0.8)$, which correspond to no, moderate, and strong pleiotropy effect, respectively. We refer the reader to Table \ref{tab:sim} for the specific values of $v$, $\rho$ and Kendall's $\tau$ considered.

We assessed the performance of the CBMAT and its competitors with respect to the parameter of interest $\eta$ (i.e. $\bm{\beta}$) in terms of the type I error and power. Under the null hypothesis, 10 000 random samples were generated as described above in all simulation scenarios. Under the alternative hypothesis, 5000 random samples were generated in all the simulation scenarios, to evaluate the power of the methods. All type I error and power results were calculated with significance threshold $\alpha = 1\%$. For CBMAT, we used the package \texttt{optim} in \texttt{R} to estimate the copula parameter as well as the parameters of the marginal distributions, under the null hypothesis of no genetic association.

{\it Analysis of simulated data:} Three simulation settings were investigated to evaluate the methods performance. Setting 1 aims to evaluate the robustness of all methods against the non-normality assumption of the marginal distributions, Setting 2 illustrates the performance of CBMAT in the case of the mixed-trait case, and Setting 3 focuses on the impact of misspecifying the true copula model for the CBMAT approach.

{\it Setting 1}: In this setting, both traits were assumed to be continuous, and we considered two cases for the marginal distributions: 1) $(F_1, F_2)$ are set to be (Exponential, Exponential) CDFs with the log-link function for the covariates, 2) $F_1$ is the Student-t distribution with $\nu = 3$ degrees of freedom and we used the identity-link function for the covariates, and $F_2$ is set to be an Exponential distribution (log-link). The dependence between the traits is generated from either a Normal, a Frank, or a Clayton copula. For CBMAT, we evaluated its performance with the correct copula, and with both the correct marginal distributions of the phenotypes (true), and when the margins are assumed to be unknown. In the latter case, selection of the marginal models was based on the best fitted GLM models in \eqref{marginalmean} with the lowest AIC, among the Gaussian, Gamma and Student-t distributions. For the other methods, as these approaches theoretically require the traits normality assumption, we compared their performance when an inverse normal transformation (INT) was applied over the residuals of each trait after adjusting for covariates (\citealp {INT}), and without normalizing the traits.

{\it Setting 2:} For the mixed binary-continuous bivariate phenotype, $Y_1$ is assumed to follow a probit latent distribution and $Y_2$ follows an Exponential distribution (log-link). In this setting, we assessed the performance of CBMAT when the fitted marginal models are the true ones, and when the marginal distribution of $Y_2$ is considered unknown, but it is selected based on the AIC criterion, in a similar way as in Setting 1. Although both MURAT and Multi-SKAT are designed to handle only continuous phenotypes, we fitted both models in order to evaluate their performance in the mixed-trait case. For both methods, we analysed the traits with and without the INT transformation over the residuals.

Of note, in both Settings 1 and 2, CBMAT was fitted with the true copula that generates the dependence structure between the traits. The impact of misspecifying the traits' dependence structure on the performance of CBMAT is given next in the Setting 3. 

{\it Setting 3:} We aimed to evaluate the robustness of CBMAT when we misspecify the dependence structure (i.e. true copula) of the two phenotypes. More specifically, we used a Normal copula to simulate either a bivariate continuous or a mixed binary-continuous phenotype and then evaluated the type I error of CBMAT when it is fitted with a Normal, a Clayton or a Frank copula to model the traits dependence. Also, to mimic analysis of real data where the phenotypes' true copula model is unknown, we evaluated the performance of CBMAT where the fitted copula was chosen following the AIC criterion. Thus, AIC criterion is derived based on the joint model (\ref{continuous-traits}) when both traits are continuous, or based on the joint model (\ref{mixed-traits}) for the mixed-trait case. The traits' marginal distributions are set to be Exponential (log-link) or Student-t ($\nu = 3$, identity link) for continuous traits and probit latent model for the binary traits.

\section{Results}
\label{sec4}
\subsection{Type I Error Rates}

{\it Setting 1:} Empirical type I error rates for continuous bivariate traits are presented in Table \ref{tab:res3}, when both traits are distributed following Exponential distributions, and in Table \ref{tab:res4} when $Y_1$ follows an Exponential distribution and $Y_2$ follows a Student-t distribution, under the null hypothesis ($\eta =0$). From these tables, one can see that both MURAT and Multi-SKAT methods have inflated type I error rates when the traits are not normally distributed (no INT), except for Multi-SKAT with the homogeneous effects assumption (Multi-SKAT Hom) in the case where $Y_2$ follows the Student-t distribution (Table \ref{tab:res4}), which shows a conservative type I error rate. Although the INT transformation seems to be effective for correcting the type I error rate of both MURAT and Multi-SKAT, the methods become relatively conservative after the traits' transformation, except for Multi-SKAT with the heterogeneous effects assumption, when both traits are distributed following Expenential distributions (Table \ref{tab:res3}). The dependence structure of the phenotypes (Normal, Frank, or Clayton) seem to have similar impact on the results of the two methods. In contrast, the type I error rate of CBMAT, in all scenarios of Setting 1, is well controlled when both correct marginal distributions (true) are used and also when the traits' margins are chosen based on the AIC criterion.

{\it Setting 2:}
Empirical type I error rates for the mixed-trait case are outlined in Table \ref{tab:res1} where $Y_1$ follows a Probit marginal model and $Y_2$ follows a marginal Exponential distribution. From this table, one can notice that Multi-SKAT, in general, shows conservative type I error rates, for both cases: with and without INT transformation of the traits. MURAT shows a type I error rate inflation when the traits are not transformed, and it becomes relatively conservative after the traits' INT. Again, in this setting, CBMAT has well controlled type I error rates, in all scenarios.

{\it Setting 3:} Empirical type I error rates for potential misspecifications of the copula model are presented in Table \ref{tab:res5}. One can notice that using the AIC criterion for selecting a copula to model the two traits dependence effectively controls type I error rates of CBMAT. On the other hand, mispecifiying a copula in the case of weakly correlated traits (Kendall's $\tau=0.05$), has no impact on the type error I rate. For the scenarios with moderate and high residual correlation, the inflation becomes quite important when the true copula is misspecified, especially for the case of using a Clayton copula instead of the (correct) Gaussian copula. This can be explained by the fact that the Clayton copula, in contrary to the Normal copula, models  asymmetric dependence with greater dependence in the negative tail than in the positive (\citealp{Cuvelier2005}).
\subsection{Power}

\begin{figure}[!p]
\centering
\includegraphics[width=1\textwidth]{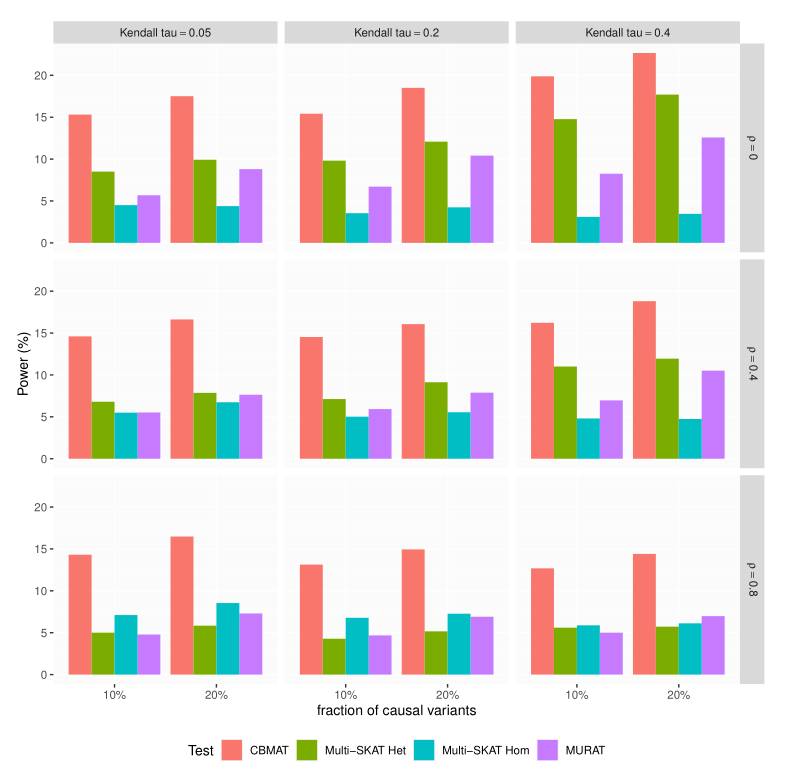}
\caption{Power $(\%)$ to detect SNPs/phenotype association. Results are computed from 5 000 data sets generated under a Gaussian copula model when either $10\%$ or $20\%$ of the variants are causal and when both traits are continuous and both follow Exponential marginal distributions. The variance component parameter $\eta$ was fixed such that the traits heritabilty under a linear model is approximately $2\%$. The compared methods are CBMAT fitted using marginal models chosen based on AIC, MURAT and Multi-SKAT fitted with INT transformation after adjusting for the covariates; Multi-SKAT is fitted with homogeneous (Multi-SKAT Hom) and uncorrelated (Multi-SKAT Het) effect sizes.}
\label{fig:power.3}
\end{figure}

\begin{figure}[!p]
\centering
\includegraphics[width=1\textwidth]{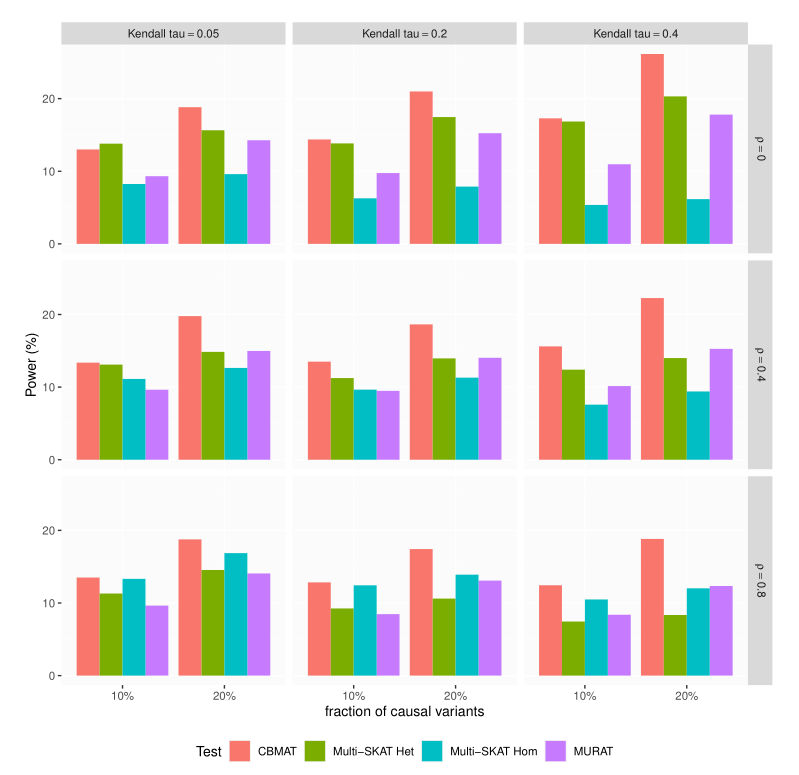}
\caption{Power $(\%)$ to detect SNPs/phenotype association. Results are computed from 5 000 data sets generated under a Gaussian copula model when either $10\%$ or $20\%$ of the variants are causal and when both traits are continuous and follow Exponential/Student-t marginal distributions. The variance component parameter $\eta$ was fixed such that the traits heritabilty under a linear model is approximately $2\%$. The compared methods are CBMAT fitted using marginal models chosen based on AIC, MURAT and Multi-SKAT fitted with INT transformation after adjusting for the covariates; Multi-SKAT is fitted with homogeneous (Multi-SKAT Hom) and uncorrelated (Multi-SKAT Het) effect sizes.}
\label{fig:power.4}
\end{figure}

\begin{figure}[!p]
\centering
\includegraphics[width=1\textwidth]{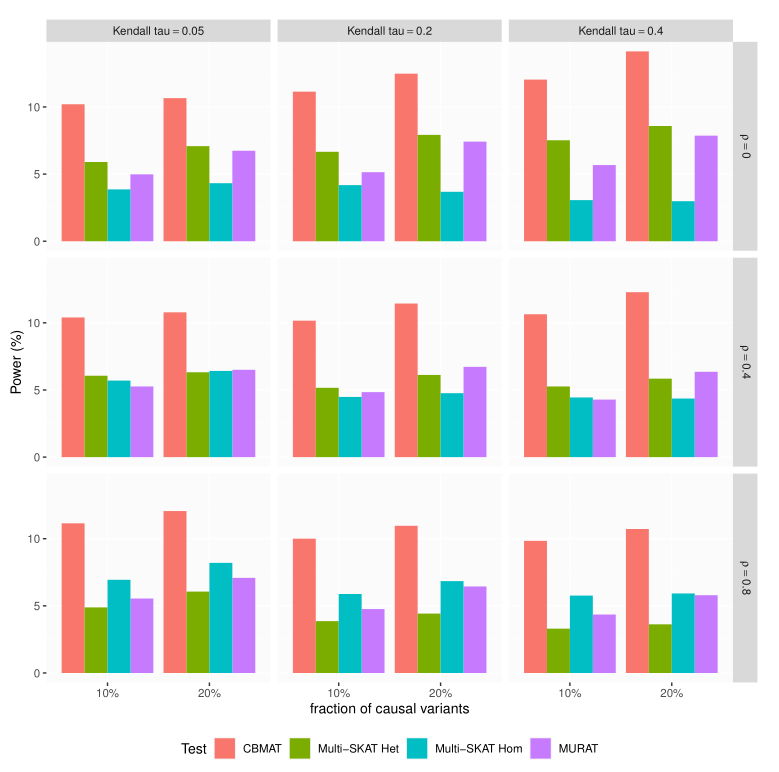}
\caption{Power $(\%)$ to detect SNPs/phenotype association. Results are computed from 5 000 data sets generated under a Gaussian copula model when either $10\%$ or $20\%$ of the variants are causal and when traits are mixed binary-continuous and follow Probit/Exponential marginal distributions. The variance component parameter $\eta$ was fixed such that the traits heritabilty under a linear model is approximately $2\%$. The compared methods are CBMAT fitted using marginal models chosen based on AIC, MURAT and Multi-SKAT fitted with INT transformation after adjusting for the covariates; Multi-SKAT is fitted with homogeneous (Multi-SKAT Hom) and uncorrelated (Multi-SKAT Het) effect sizes.}
\label{fig:power.1}
\end{figure}

{\it Settings 1 and 2:}
We show the empirical power results for the continuous-traits case in Figures \ref{fig:power.3} and  \ref{fig:power.4}, and for the mixed-traits case in Figure \ref{fig:power.1}. The traits dependence structure is generated from the Gaussian copula. Empirical results, when the dependence structure is generated from a Clayton copula are presented in Appendix E of the Supplementary Materials. For the CBMAT approach, results were reported when the marginal distributions are selected based on the best fitted marginal models using the AIC criterion. For MURAT and Multi-SKAT, we reported their empirical power after INT transformation of the residuals since both methods have shown inflated type I error rates when the traits are not normally distributed.

From these figures, one can notice that CBMAT has greater power compared to MURAT and Multi-SKAT for almost all scenarios. Moreover, CBMAT power to detect association is not affected when the margins are chosen based on the AIC criterion compared to when CBMAT is fitted with the true/correct marginal distributions, as shown in Supplementary Tables of Appendix E.

\subsection{ALSPAC data analysis}

We illustrate our method on the \textit {Avon Longitudinal Study of Parents and Children} (ALSPAC) data. This transgenerational prospective study is concerned with the genetic, epigenetic, biological, psychological and social factors influencing various indicators of health and social development over a lifetime (\citealp{ALSPAC}). The cohort is composed of children born between 1990 and 1992 in the Bristol area in the United Kingdom. For analysis, we focused on 1477 subjects for whom the entire genome has been sequenced. In addition, several clinical phenotypes have been measured for these subjects, including high-density lipoprotein (HDL) and low-density lipoprotein (LDL), responsible for cholesterol transport to the liver, and triglycerides (Trigl), apolipoproteins B (ApoB) and apolipoproteins A1 (ApoA1), which are respectively the main constituents of low and high density lipoproteins. We are interested in the association between these phenotypes and two sets of SNPs falling within the {\it APOC3} and {\it APOA1} genes, located on chromosome 11. 
Previous studies have shown that mutations in the {\it APOA1} gene cause a decrease in HDL levels in the blood, potentially increasing the risk of cardiovascular disease (\citealp{APOA1}) while mutations in the {\it APOC3} gene are associated with low plasma triglyceride levels and reduced risk of ischemic cardiovascular disease (\citealp{APOC3}.

Pearson correlation coefficient values for each pair of phenotypes are presented in Appendix E of the Supplementary Materials. As expected, the correlation between HDL and ApoA1 (0.828) is very high. In addition, HDL is also strongly correlated with Trigl. Thus, the pairs of phenotypes (HDL, ApoA1) and (HDL, Trigl) are considered for association analysis. Box plots for HDL, Trigl, and ApoA1 measurements before and after logarithmic transformation are presented in Appendix F of the Supplementary Materials. Thus, assuming ApoA1 and Trigl traits follow log-normal distributions, we normalized them using a logarithmic transformation. In order to mimic the mixed-trait case of the analysis of this data, 
we dichotomized HDL using the sample median as a cutoff value to create a binary trait. For MURAT and Multi-SKAT, we first fitted a linear regression model of the HDL binary trait and adjusted for Sex as a covariate, before applying an INT transformation of the residuals.

The CBMAT approach is fitted directly to the traits without transformations as the method handles both dichotomous and non-normally-distributed traits. More precisely, we selected marginal models of the continuous traits, ApoA1 and Trig, among Gaussian, Gamma and Student-t distributions, based on AIC criterion. This procedure suggests Student-t distributions with 12 and 3 degrees of freedom as best marginal models for ApoA1 and Trigl, respectively. The choice of the copula to model the traits dependence structure is also conducted based on AIC criterion under the null hypothesis. The Frank and Gaussian copulas are selected as the best joint models for (HDL, ApoA1) and (HDL, Trigl), respectively. The estimated copula parameters, the corresponding Kendall's $\tau$, and pleiotropy coefficient for each bivariate phenotype are presented in Table \ref{tableau ALSPAC.2}.

The two sets of SNPs that we analysed consist of 27 SNPs located in the {\it APOA1} gene (MAF ranging from 0.0003 to 0.45), and 59 SNPs located in the {\it APOC3} gene (MAF ranging from 0.0003 to 0.3). The p-values for the Multi-SKAT, MURAT and CBMAT approaches are presented in Table \ref{tableau ALSPAC.3}. As observed in the simulations, the power of Multi-SKAT decreases more importantly when the pleiotropy is low and homogeneous effect sizes on different phenotypes are assumed. This may explain why the association p-values for (HDL,Trigl) are significantly higher than for (HDL,ApoA1). For all four association tests, CBMAT is more powerful than both both MURAT and Multi-SKAT, except for the analysis of (HDL, ApoA1), where Multi-SKAT has smaller p-values. However, Multi-SKAT results need to be interpreted with caution as the method has shown inflation and/or power loss in our simulation studies. Finally, recall that, as opposed to the other methods, CBMAT allows to test for association and provides covariates' effects estimation directly on the original phenotypes, without any transformation.

\section{Discussion}
\label{sec5}
In this study, we have introduced a new copula based method for rare‐variant association tests for both continuous and mixed bivariate phenotypes, where binary traits are incorporated through the use of a marginal latent probit model. As demonstrated, CBMAT is more flexible with regard to modeling the dependence structure between the traits through the use of copulas. Moreover, our proposed test is applicable for a wide range of distributions as it does not assume normality of the traits' margins. The simulation studies show that the type I error of our proposed data-driven score-type test is well controlled for both continuous and mixed bivariate phenotypes, even when both the true copula and marginal distributions were not known. The simulation studies investigation also shows that methods which assume multivariate normality, such as MURAT and Multi-SKAT, are prone to type I error inflation and a loss of power to detect association when applied to non-normally distributed outcomes. 

Our method is a generalization of MURAT when bivariate normality is not assumed. Hence, when one uses normal marginal distributions and a Gaussian copula to model the joint distribution of the traits, CBMAT reduces to MURAT. One of the main advantages of our method, though, is that we model the dependence structure between traits regardless of their marginal distributions. Moreover, the choice among different copula families to model the traits dependence allows for more modeling flexibility. Furthermore, our method circumvent inherent complications in terms of analysis and interpretation resulting from transformation of variables prior to analysis. Indeed, transformation of the original data may not always result in accurate interpretation of the estimated coefficients for the covariates and/or the genetic effect, especially since several clinical measurements can only be modeled by distributions taking positive values. 
Our method is also applicable for studies where there is clinical utility in dichotomizing a trait or when one outcome consists of disease status, and we aim to analyse such a trait with other related continuous phenotypes. Most existing region-based genetic association tests do not have current implementations for the mixture of binary-continuous traits.

We assumed a bivariate phenotype throughout the paper; however, in many applications, researchers have access to more than two traits. Although the mathematical derivation of the CBMAT approach in the case where $K > 2$ is straightforward, work still needs to be made for the method implementation and its computational efficiency in order to scale to GWAS data sets. Also, Archimedean copulas in this case, such as Clayton and Frank copulas, might not be suitable anymore to model dependence between the traits as these two copula families only allow for exchangeable dependence structure between the $K$ phenotypes. Finally, another useful extension of our methodology would be to take advantage of the great flexibility provided by copulas to allow for several forms of clustering, such as in family-based or longitudinal studies. For instance, in famliy-based designs, one needs to capture both between- and within-subjects dependencies. Thus, again marginal GLM models could link the traits to the covariates and the genetic region, and Vine copulas (\citealp{joe2014dependence}) could be exploited to model the additional within-family source of dependence and develop a copula-based extension of CBMAT for related subjects.

\section{Software}
\label{sec6}

A software R package, together with a toy data example for illustration and complete documentation, are available at \url{https://github.com/julstpierre/CBMAT}.

\section*{Acknowledgments}
This research supported by both the Fonds de recherche Québec-Santé through individual grant $\#$ 267074 and by the Natural Sciences and Engineering Research Council of Canada through individual discovery research grants to Karim Oualkacha. 
This study makes use of data generated by the UK10 K Consortium, derived from samples from ALSPAC, under data access agreement ID2250. A full list of the investigators who contributed to the generation of the data is available from \url{www.UK10K.org}. Funding for UK10 K was provided by the Wellcome Trust under award {\bf WT091310}.

\section*{Data availability statement}
Data is available under request from UK10K consortium or from ALSPAC consortium. We have access to this data through UK10K consortium since Dr Oualkacha is a member of the statistics group of the UK10K consortium.

{\it Conflict of Interest}: None declared.

\bibliographystyle{biorefs}
\bibliography{biblio}

\begin{table}[!p]
\centering
\caption{Values of $v$, $\rho$, $\tau$ in the simulations (leading to 21 distinct scenarios). $H_0$ and $H_a$ stand for the null and the alternative hypotheses of the association between the genetic region and the two phenotypes.
$^a$ fraction of variants that are causal and induce the association between the genetic region and the two traits; $^b$ region-specific pleiotropy effect; $^c$ Kendall's $\tau$ in the CBMAT copula model.} 
\label{tab:sim}
\begin{tabular}{llll}
\hline
      & $v^a$         & $\rho^b$        & Kendall's $\tau^c$     \\ \hline
$H_0$ &           0      & 0               & \{0.05, 0.20, 0.40\} \\ \hline
$H_a$ & \{10\%, 20\%\}   & \{0, 0.4, 0.8\} & \{0.05, 0.20, 0.40\} \\ \hline
\end{tabular}
\end{table}

\begin{table}[!p]
\centering
\caption{
Empirical type 1 error rate $(\%)$ under the null hypothesis of no SNPs/phenotype association $(\eta = 0)$ where the data are generated under either Gaussian, Frank or Clayton copula models. Results are computed from 10 000 data sets generated under Setting 1 where both traits are continuous and both follow Exponential marginal distributions. The compared methods are CBMAT fitted with true marginal models (true), and fitted using marginal models chosen based on AIC (AIC);  MURAT and Multi-SKAT are fitted without INT transformation (no INT) and with INT transformation after adjusting for the covariates (INT); Multi-SKAT is fitted with homogeneous (Multi-SKAT Hom) and uncorrelated (Multi-SKAT Het) effect sizes.}
\label{tab:res3}
\begin{tabular}{cccccccccc}
  \hline & & \multicolumn{2}{c}{CBMAT} & \multicolumn{2}{c}{MURAT}  & \multicolumn{2}{c}{Multi-SKAT Het} & \multicolumn{2}{c}{Multi-SKAT Hom} \\
 copula & Kendall's $\tau$ & true & AIC & no INT & INT & no INT & INT & no INT & INT \\
  \hline
Normal & 0.05 & 1.11 & 1.11 & 2.34 & 0.84 & 4.10 & 1.08 & 4.08 & 0.60 \\ 
   & 0.20 & 1.19 & 1.18 & 2.27 & 0.84 & 4.13 & 1.18 & 4.10 & 0.59 \\ 
   & 0.40 & 1.19 & 1.19 & 2.27 & 0.90 & 4.12 & 1.16 & 4.10 & 0.45 \\ 
   \hline
Franck & 0.05 & 1.14 & 1.14 & 2.26 & 0.80 & 4.21 & 1.10 & 4.19 & 0.58 \\ 
   & 0.20 & 1.22 & 1.22 & 2.31 & 0.80 & 4.24 & 1.14 & 4.18 & 0.58 \\ 
   & 0.40 & 1.14 & 1.14 & 2.31 & 0.89 & 4.31 & 1.18 & 4.26 & 0.48 \\ 
   \hline
Clayton & 0.05 & 1.06 & 1.06 & 2.33 & 0.76 & 4.15 & 1.09 & 4.16 & 0.59 \\ 
   & 0.20 & 1.18 & 1.18 & 2.30 & 0.88 & 4.12 & 1.10 & 4.12 & 0.59 \\ 
   & 0.40 & 1.06 & 1.06 & 2.26 & 0.82 & 4.09 & 1.05 & 4.07 & 0.56 \\ 
   \hline
\end{tabular}
\end{table}

\begin{table}[!p]
\centering
\caption{Empirical type 1 error rate $(\%)$ under the null hypothesis of no SNPs/phenotype association $(\eta = 0)$ where the data are generated under either Gaussian, Frank or Clayton copula models. Results are computed from 10 000 data sets generated under Setting 1 where both traits are continuous and follow Exponential/Student-t marginal distributions. The compared methods are CBMAT fitted with true marginal models (true), and fitted using marginal models chosen based on AIC (AIC);  MURAT and Multi-SKAT are fitted without INT transformation (no INT) and with INT transformation after adjusting for the covariates (INT); Multi-SKAT is fitted with homogeneous (Multi-SKAT Hom) and uncorrelated (Multi-SKAT Het) effect sizes.}
\label{tab:res4}
\begin{tabular}{cccccccccc}
  \hline & & \multicolumn{2}{c}{CBMAT} & \multicolumn{2}{c}{MURAT}  & \multicolumn{2}{c}{Multi-SKAT Het} & \multicolumn{2}{c}{Multi-SKAT Hom} \\
 copula & Kendall's $\tau$ & true & AIC & no INT & INT & no INT & INT & no INT & INT \\
  \hline
Normal & 0.05 & 0.93 & 0.96 & 1.57 & 0.93 & 3.50 & 0.76 & 0.02 & 0.77 \\ 
   & 0.20 & 1.00 & 1.02 & 1.35 & 0.91 & 3.95 & 0.67 & 0.01 & 0.89 \\ 
   & 0.40 & 1.07 & 1.10 & 1.37 & 0.89 & 4.48 & 0.86 & 0.02 & 0.85 \\ 
   \hline
Franck & 0.05 & 0.92 & 0.95 & 1.61 & 0.90 & 3.46 & 0.75 & 0.02 & 0.79 \\ 
   & 0.20 & 0.91 & 0.93 & 1.48 & 0.91 & 3.94 & 0.73 & 0.02 & 0.88 \\ 
   & 0.40 & 0.98 & 1.04 & 1.53 & 0.95 & 4.50 & 0.88 & 0.02 & 0.88 \\ 
   \hline
Clayton & 0.05 & 0.86 & 0.90 & 1.43 & 0.89 & 3.54 & 0.80 & 0.02 & 0.81 \\ 
   & 0.20 & 0.95 & 0.99 & 1.36 & 0.96 & 3.90 & 0.69 & 0.02 & 0.80 \\ 
   & 0.40 & 1.01 & 1.03 & 1.37 & 0.83 & 4.36 & 0.83 & 0.03 & 0.85 \\ 
   \hline
\end{tabular}
\end{table}

\begin{table}[!p]
\centering
\caption{Empirical type 1 error rate $(\%)$ under the null hypothesis of no SNPs/phenotype association $(\eta = 0)$ where the data are generated under either Gaussian, Frank or Clayton copula models. Results are computed from 10 000 data sets generated under Setting 2 for mixed-trait case where $Y_1$ follows a Probit marginal model and $Y_2$ follows a marginal Exponential distribution. The compared methods are CBMAT fitted with true marginal models (true), and fitted using marginal models chosen based on AIC (AIC);  MURAT and MUlti-SKAT are fitted without INT transformation (no INT) and with INT transformation after adjusting for the covariates (INT); Multi-SKAT is fitted with homogeneous (Multi-SKAT Hom) and uncorrelated (Multi-SKAT Het) effect sizes.}
\label{tab:res1}
\begin{tabular}{cccccccccc}
  \hline & & \multicolumn{2}{c}{CBMAT} & \multicolumn{2}{c}{MURAT}  & \multicolumn{2}{c}{Multi-SKAT Het} & \multicolumn{2}{c}{Multi-SKAT Hom} \\
 copula & Kendall's $\tau$ & true & AIC & no INT & INT & no INT & INT & no INT & INT \\
  \hline
Normal & 0.05 & 1.14 & 1.14 & 2.17 & 0.78 & 0.03 & 0.40 & 0.03 & 0.64 \\ 
   & 0.20 & 1.15 & 1.15 & 2.19 & 0.81 & 0.03 & 0.40 & 0.02 & 0.55 \\ 
   & 0.40 & 1.19 & 1.19 & 2.13 & 0.82 & 0.02 & 0.44 & 0.02 & 0.62 \\ 
   \hline
Franck & 0.05 & 1.10 & 1.10 & 2.16 & 0.78 & 0.03 & 0.40 & 0.03 & 0.63 \\ 
   & 0.20 & 1.11 & 1.11 & 2.20 & 0.83 & 0.02 & 0.37 & 0.02 & 0.55 \\ 
   & 0.40 & 1.05 & 1.05 & 2.11 & 0.79 & 0.02 & 0.43 & 0.02 & 0.61 \\ 
   \hline
Clayton & 0.05 & 1.12 & 1.12 & 2.18 & 0.82 & 0.03 & 0.43 & 0.03 & 0.61 \\ 
   & 0.20 & 1.08 & 1.08 & 2.10 & 0.78 & 0.02 & 0.42 & 0.02 & 0.60 \\ 
   & 0.40 & 1.06 & 1.06 & 2.18 & 0.80 & 0.01 & 0.37 & 0.01 & 0.69 \\ 
   \hline
\end{tabular}
\end{table}


\begin{table}[!p]
\centering
\caption{Empirical type 1 error rate $(\%)$ under the null hypothesis of no SNPs/phenotype association $(\eta = 0)$. Results are computed from 10 000 data sets generated under Setting 3 where we used a Normal copula to simulate either a bivariate continuous or a mixed binary-continuous phenotype and then fitted CBMAT with a Normal, Clayton or Frank copula, or a model where the copula was chosen based on AIC.}
\label{tab:res5}
\begin{tabular}{llccccc}
  \hline & & &\multicolumn{4}{c}{Copula} \\
 $F_1$ & $F_2$ & Kendall's $\tau$ & Normal & Clayton & Frank & AIC\\
  \hline
Binomial & Exponential & 0.05 & 1.00 & 1.03 & 1.08 & 1.05 \\ 
   &  & 0.20 & 1.07 & 1.24 & 1.14 & 1.08 \\ 
   &  & 0.40 & 1.12 & 1.44 & 1.31 & 1.09 \\ 
   \hline
Exponential & Exponential & 0.05 & 1.31 & 1.32 & 1.29 & 1.27 \\ 
   &  & 0.20 & 1.34 & 1.53 & 1.42 & 1.34 \\ 
   &  & 0.40 & 1.26 & 2.10 & 1.77 & 1.26 \\ 
   \hline
 & Student & 0.05 & 1.07 & 1.19 & 1.11 & 1.11 \\ 
   &  & 0.20 & 1.13 & 1.47 & 1.25 & 1.24 \\ 
   &  & 0.40 & 1.12 & 1.66 & 1.52 & 1.51 \\ 
   \hline
\end{tabular}
\end{table}


\begin{table}[!p]
\centering
\caption{Estimated copula parameter $\hat\theta$, corresponding Kendall's $\hat\tau$ and pleitropic coefficient $\hat\rho$ for each mixed bivariate phenotype.}
\label{tableau ALSPAC.2}
\begin{tabular}{lllll}
\hline
Phenotype  & Copula   & $\hat\theta$ & Kendall's $\hat\tau$ & $\hat\rho$ \\ \hline
(HDL,ApoA1) & Frank   & 7.05& 0.56          & 0.4    \\
(HDL,Trigl) & Gaussian & -0.47 & -0.31         & 0      \\ \hline
\end{tabular}
\end{table}

\begin{table}[!p]
\centering
\caption{P-values for Multi-SKAT, MURAT and CBMAT under the null hypothesis of no gene/phenotype association for APOA1 and APOC3 genes. HDL trait has been dichotomized, while ApoA1 and Trigl traits have been log-transformed prior to analysis. All association tests have been adjusted for sex as a potential confounder.}
\label{tableau ALSPAC.3}
\begin{tabular}{lcccc}
\hline       
Gene    & \multicolumn{4}{c}{}  \\ 
\ \ Phenotype & Multi-SKAT Het  & Multi-SKAT Hom   		& MURAT  				& CBMAT \\ \hline
APOA1 & & & & \\
\ - (HDL,ApoA1)  & $2.44\times10^{-3}$ & $9.40\times10^{-4}$ & $5.38\times10^{-2}$ & $5.99\times10^{-3}$ \\ 							 
\ - (HDL,Trigl) & $1.10\times10^{-2}$ & $1.77\times10^{-2}$ & $4.49\times10^{-4}$ & $6.82\times10^{-5}$ \\  

& & & & \\

APOC3 & & & & \\
\ - (HDL,ApoA1) & $1.99\times10^{-3}$& $6.15\times10^{-4}$ & $9.36\times10^{-3}$ & $1.86\times10^{-4}$  \\         									 
\ - (HDL,Trigl) & $2.61\times10^{-2}$ & $2.50\times10^{-2}$ & $6.00\times10^{-4}$& $2.29\times10^{-4}$  \\  

\hline

\end{tabular}
\end{table}

\end{document}


\title{Supplementary Materials to copula-based set-variant association test for bivariate continuous or mixed phenotypes}

\author{JULIEN ST-PIERRE$^{\ast}$\\[4pt]
\textit{Department of Epidemiology, Biostatistics and Occupational Health, McGill University, Montréal, Québec, Canada}
\\[2pt]
{julien.st-pierre@mail.mcgill.ca}\\[4pt]
KARIM OUALKACHA\\[4pt]
\textit{Département de Mathématiques, Université du Québec à Montréal, Montréal, Québec, Canada}}

\markboth%
{J. St-Pierre and K. Oualkacha}
{Supplementary to copula-based association test for bivariate mixed phenotypes}

\maketitle

\footnotetext{To whom correspondence should be addressed.}
\section*{Appendix A. Derived compact form of the variance component score test $U(\eta)$}

The conditional log-likelihood function when both phenotypes are continuous is equal to
\begin{align}
\label{condloglik}
l(\bm{\xi}|\bm{\beta})&=\sum_{i=1}^{n}\left[\text{log }f_1(y_{i1}|\bm{\beta})+\text{log }f_2(y_{i2}|\bm{\beta})+\text{log }c_{\theta}(F_1(y_{i1}|\bm{\beta});F_2(y_{i2}|\bm{\beta}))\right]
\end{align}
where $c_\theta(F_1(y_{i1}|\bm{\beta});F_2(y_{i2}|\bm{\beta}))$ is the density of the copula and $\theta$ is a dependence parameter measuring the dependence between marginal CDFs.
Assuming marginal densities come from exponential families, we have, for $j=1,2$,
\begin{align*}
\frac{\partial}{\partial\bm{\beta}}\text{log }f(y_{ij}|\bm{\beta})&=\frac{\partial}{\partial\bm{\beta}}[\frac{y_{ij}\xi_{ij}-b(\xi_{ij})}{\phi_j}-c(y_{ij},\phi_j)]\\
&=\frac{1}{\phi_j}[y_{ij}\frac{\partial \xi_{ij}}{\partial\bm{\beta}}-\frac{\partial b(\xi_{ij})}{\partial \xi_{ij}}\frac{\partial \xi_{ij}}{\partial\bm{\beta}}]\nonumber\\
&=\frac{1}{\phi_j}[y_{ij}-\mu_{ij}]\frac{\partial \xi_{ij}}{\partial\bm{\beta}}.
\end{align*}
We simplify using chain rule
\begin{align*}
\frac{\partial}{\partial\bm{\beta}}\text{log }f(y_{ij}|\bm{\beta})&=\frac{1}{\phi_j}[y_{ij}-\mu_{ij}]\frac{\partial \xi_{ij}}{\partial \mu_{ij}}\frac{\partial \mu_{ij}}{\partial\bm{\beta}}\nonumber\\
&=\frac{1}{\phi_j}[y_{ij}-\mu_{ij}]\frac{\partial \mu_{ij}}{\partial \xi_{ij}}^{-1}\frac{\partial \mu_{ij}}{\partial g(\mu_{ij})}\frac{\partial g(\mu_{ij})}{\partial\bm{\beta}}\nonumber\\
&=\frac{1}{\phi_j}[y_{ij}-\mu_{ij}]\frac{\partial^2 b(\xi_{ij})}{\partial \xi_{ij}^2}^{-1}\frac{\partial g(\mu_{ij})}{\partial \mu_{ij}}^{-1}\frac{\partial g(\mu_{ij})}{\partial\bm{\beta}}\nonumber.
\end{align*}
We define the variance function $v(\mu)=b''(\xi)$, from where
\begin{align*}
\frac{\partial}{\partial\bm{\beta}}\text{log }f(y_{ij}|\bm{\beta})&=\frac{1}{\phi_j}[y_{ij}-\mu_{ij}]v(\mu_{ij})^{-1}\frac{\partial g(\mu_{ij})}{\partial \mu_{ij}}^{-1}\begin{bmatrix}\frac{\partial g(\mu_{ij})}{\partial\bm{\beta}_1}\\[0.8em]\frac{\partial g(\mu_{ij})}{\partial\bm{\beta}_2}\end{bmatrix}\nonumber\\
&=\frac{1}{\phi_j}\frac{[y_{ij}-\mu_{ij}]}{v(\mu_{ij})\frac{\partial g(\mu_{ij})}{\partial \mu_{ij}}}\begin{bmatrix}\mathbbm{1}_{1}(j) \\[0.8em] \mathbbm{1}_{2}(j) \end{bmatrix}\otimes\bm{G}_i
\end{align*}
with $G_i\in\mathbbm{R}^r$ a column vector containing the $r$ SNPs for the $i^{th}$ individual and $$\mathbbm{1}_{1}(j)=\begin{cases}1 & \text{, if }j=1 \\ 0 & \text{otherwise} \end{cases}.$$
Therefore, the gradient of the conditional log-likelihood with respect to $\bm{\beta}$ is given by
\begin{align*}
\frac{\partial}{\partial\bm{\beta}}l(\bm{\xi|\beta})
&=\sum_{i=1}^{n}\begin{bmatrix}\frac{y_{i1}-\mu_{i1}}{\phi_1v(\mu_{i1})\frac{\partial g(\mu_{i1})}{\partial \mu_{i1}}}\\[1.5em]\frac{y_{i2}-\mu_{i2}}{\phi_2v(\mu_{i2})\frac{\partial g(\mu_{i2})}{\partial \mu_{i2}}}\end{bmatrix}\otimes\bm{G}_i+\sum_{i=1}^{n}\frac{\partial}{\partial\bm{\beta}}\text{log}\ c_{\alpha}\nonumber\\[0.8em]
&=\sum_{i=1}^{n}\begin{bmatrix}\frac{y_{i1}-\mu_{i1}}{\phi_1v(\mu_{i1})\frac{\partial g(\mu_{i1})}{\partial \mu_{i1}}}\\[1.5em]\frac{y_{i2}-\mu_{i2}}{\phi_2v(\mu_{i2})\frac{\partial g(\mu_{i2})}{\partial \mu_{i2}}}\end{bmatrix}\otimes\bm{G}_i+\sum_{i=1}^{n}\begin{bmatrix}\frac{\partial}{\partial\bm{\beta}_1}\text{log}\ c_{\alpha}\\[0.8em] \frac{\partial}{\partial\bm{\beta}_2}\text{log}\ c_{\alpha}\end{bmatrix} \nonumber\\[0.8em]
&=\sum_{i=1}^{n}\begin{bmatrix}\frac{y_{i1}-\mu_{i1}}{\phi_1v(\mu_{i1})\frac{\partial g(\mu_{i1})}{\partial \mu_{i1}}}\\[1.5em]\frac{y_{i2}-\mu_{i2}}{\phi_2v(\mu_{i2})\frac{\partial g(\mu_{i2})}{\partial \mu_{i2}}}\end{bmatrix}\otimes\bm{G}_i+\sum_{i=1}^{n}\begin{bmatrix}\frac{\partial\text{log}(c_{\alpha})}{\partial \mu_{i1}}\frac{\partial \mu_{i1}}{\partial\bm{\beta}_1}\\[0.8em] \frac{\partial\text{log}(c_{\alpha})}{\partial \mu_{i2}}\frac{\partial \mu_{i2}}{\partial\bm{\beta}_2}\end{bmatrix} \nonumber\\[0.8em]
&=\sum_{i=1}^{n}\begin{bmatrix}\frac{y_{i1}-\mu_{i1}}{\phi_1v(\mu_{i1})\frac{\partial g(\mu_{i1})}{\partial \mu_{i1}}}+\frac{\frac{\partial\text{log}(c_{\alpha})}{\partial \mu_{i1}}}{\frac{\partial g(\mu_{i1})}{\partial \mu_{i1}}}\\[1.5em]\frac{y_{i2}-\mu_{i2}}{\phi_2v(\mu_{i2})\frac{\partial g(\mu_{i2})}{\partial \mu_{i2}}}+\frac{\frac{\partial \text{log}(c_{\alpha})}{\partial \mu_{i2}}}{\frac{\partial g(\mu_{i2})}{\partial \mu_{i2}}}\end{bmatrix}\otimes\bm{G}_i.
\end{align*}
Using a more compact form, we can rewrite the previous equation as
\begin{align}
\label{eq:compact1}
\frac{\partial}{\partial\bm{\beta}}l(\bm{\xi|\beta}) &= \sum_{i=1}^n\begin{bmatrix} L_{1i} \\[1em] L_{2i} \end{bmatrix} \otimes\bm{G}_i \nonumber \\
&= \begin{bmatrix} \bm{G}^T & \bm0 \\[1em] \bm0 & \bm{G}^T \end{bmatrix} \begin{bmatrix} \bm L_1 \\[1em] \bm L_2 \end{bmatrix} \nonumber \\
&= \left(I_2\otimes \bm{G}^T\right) \bm{L},
\end{align}
where $\bm{G}$ is the $n\times r$ genotypes matrix for all individuals. The Hessian of the conditional log-likelihood with respect to $\bm{\beta}$ is given by
\begin{align*}
\frac{\partial^2}{\partial\bm{\beta}\partial\bm{\beta}^T}l(\bm{\xi|\beta})&=\sum_{i=1}^{n}\begin{bmatrix}\frac{\partial}{\partial\bm{\beta_1}}[\frac{1}{\phi}\frac{y_{i1}-\mu_{i1}}{v(\mu_{i1})g'_{\mu_{i1}}}+\frac{\frac{\partial\text{log}(c_{\alpha})}{\partial\mu_{i1}}}{g'_{\mu_{i1}}}]& \frac{\partial}{\partial\bm{\beta}_1}[\frac{1}{\phi}\frac{y_{i2}-\mu_{i2}}{v(y_{i2})g'_{\mu_{i2}}}+\frac{\frac{\partial\text{log}(c_{\alpha})}{\partial\mu{i2}}}{g'_{\mu_{i2}}}]\\[0.8em] \frac{\partial}{\partial\bm{\beta}_2}[\frac{1}{\phi}\frac{y_{i1}-\mu_{i1}}{v(\mu_{i1})g'_{\mu_{i1}}}+\frac{\frac{\partial\text{log}(c_{\alpha})}{\partial\mu_{i1}}}{g'_{\mu_{i1}}}]& \frac{\partial}{\partial\bm{\beta}_2}[\frac{1}{\phi}\frac{y_{i2}-\mu_{i2}}{v(\mu_{i2})g'_{\mu_{i2}}}+\frac{\frac{\partial\text{log}(c_{\alpha})}{\partial\mu_{i2}}}{g'_{\mu_{i2}}}] \end{bmatrix}^T\otimes\bm{G}_i^T \nonumber\\[0.8em]
&=\sum_{i=1}^{n}\begin{bmatrix} \frac{\partial}{\partial\mu_{i1}}[\frac{1}{\phi}\frac{y_{i1}-\mu_{i1}}{v(\mu_{i1})g'_{\mu_{i1}}}+\frac{\frac{\partial\text{log}(c_{\alpha})}{\partial\mu_{i1}}}{g'_{\mu_{i1}}}]\frac{1}{g'_{\mu_{i1}}}& \frac{\partial}{\partial\mu_{i1}}[\frac{\frac{\partial\text{log}(c_{\alpha})}{\partial\mu_{i2}}}{g'_{\mu_{i2}}}]\frac{1}{g'_{\mu_{i1}}}\\[0.8em] \frac{\partial}{\partial\mu_{i2}}[\frac{\frac{\partial\text{log}(c_{\alpha})}{\partial\mu_{i1}}}{g'_{\mu_{i1}}}]\frac{1}{g'_{\mu_{i2}}}& \frac{\partial}{\partial\mu_{i2}}[\frac{1}{\phi}\frac{y_{i2}-\mu_{i2}}{v(\mu_{i2})g'_{\mu_{i2}}}+\frac{\frac{\partial\text{log}(c_{\alpha})}{\partial\mu_{i2}}}{g'_{\mu_{i2}}}]\frac{1}{g'_{\mu_{i2}}} \end{bmatrix}\otimes\bm{G}_i\bm{G}_i^T.
\end{align*}
Using again a more compact form, we can rewrite
\begin{align}
\label{eq:compact2}
\frac{\partial^2}{\partial\bm{\beta}\partial\bm{\beta}^T}l(\bm{\xi|\beta})&= \left(I_2\otimes \bm{G}^T\right)\frac{\partial}{\partial\beta^T}\bm{L} \nonumber\\
&= \left(I_2\otimes \bm{G}^T\right)\begin{bmatrix} \frac{\partial}{\partial\beta_1^T}\bm{L} & \frac{\partial}{\partial\beta_2^T}\bm{L} \end{bmatrix} \nonumber\\
&= \left(I_2\otimes \bm{G}^T\right) \bm D \left(I_2\otimes \bm{G}\right),
\end{align}
where $\bm{D}=\begin{bmatrix} A_{11} & A_{12} \\ A_{21} & A_{22}\end{bmatrix}$ is a block matrix, and $A_{ij}=\text{diag}\left(\partial \bm{L}_i/\partial g(\mu_j)\right)$ for $i=1,2$ and $j=1,2$.

Of note, we could derive the same formulae as in \eqref{eq:compact1} and \eqref{eq:compact2} for the mixed binary-continuous case, only the explicit form of the conditional log-likelihood in \eqref{condloglik} would be different.

Recall that the full likelihood is given by the $2r$-dimensional integral
\begin{align}
\label{eq. method.13}
L(\eta, \bm{\xi})=\int_{\bm{\beta}}L(\bm{\xi}|\bm{\beta}) \ \bm{H}(\bm{\beta}) d\bm{\beta},
\end{align}
and that the score test for $H_0$: $\eta=0$ is based on the score statistic
\begin{align}
\label{eq:score}
U(\eta)=\frac{\partial}{\partial\eta}\text{log }L(\eta,\bm{\xi}).
\end{align}
The problem with the direct computation of this score is the evaluation of the integral in dimension $\mathbbm{R}^{2r}$ in equation \eqref{eq. method.13}.
To solve this computational problem, we choose an approximation of this integral using Taylor's expansion techniques of $L(\bm{\xi}|\bm{\beta}) $ in the neighborhood of $\bm{\beta}=\bm{0}_{2r}$ (\citealp{Taylor}), that is,
\begin{equation*}
L(\bm\xi|\bm{\beta})
\approx\text{exp}\{l(\bm{\xi}|\bm{\beta})\}_{|\bm{\beta}=0}+\frac{\partial}{\partial\bm{\beta}}\left[\text{exp}\{l(\bm{\xi}|\bm{\beta})\}\right]^T_{|\bm{\beta}=\bm{0}}\cdot\bm{\beta}+\frac{1}{2}\bm{\beta}^T\frac{\partial^2}{\partial\bm{\beta}\partial\bm{\beta}^T}[\text{exp}\{l(\bm{\xi}|\bm{\beta})\}]_{|\bm{\beta}=\bm{0}}\cdot\bm{\beta}.
\end{equation*}
Using chain rule, we can write 
\begin{align}
L(\bm\xi|\bm{\beta})&\approx L(\bm{\xi}|\bm{\beta})_{|\bm{\beta}=\bm{0}}+\left[\text{exp}\{l(\bm{\xi}|\bm{\beta})\}\cdot\frac{\partial}{\partial\beta}l(\bm{\xi}|\bm{\beta})\right]^T_{|\bm{\beta}=\bm{0}}\cdot\bm{\beta} +\frac{1}{2}\bm{\beta}^T\frac{\partial}{\bm{\beta}}\left[\text{exp}\{l(\bm{\xi}|\bm{\beta})\}\cdot\frac{\partial}{\partial\bm{\beta}}l(\bm{\xi}|\bm{\beta})\right]_{|\bm{\beta}=\bm{0}}\cdot\bm{\beta}\nonumber\\
&=\left.L(\bm{\xi}|\bm{\beta})_{|\bm{\beta}=\bm{0}}+L(\bm{\xi}|\bm{\beta})_{|\bm{\beta}=\bm{0}}\cdot\frac{\partial}{\partial\bm{\beta}}l(\bm{\xi}|\bm{\beta})^T_{|\bm{\beta}=\bm{0}}\cdot\bm{\beta}\nonumber\right.\\
&\phantom{\left.{}\approx L(\bm{\xi}|\bm{\beta})_{|\bm{\beta}=\bm{0}}\right.}+\frac{1}{2}\bm{\beta}^T\left[\text{exp}\{l(\bm{\xi}|\bm{\beta})\}\left(\frac{\partial}{\partial\bm{\beta}}l(\bm{\xi}|\bm{\beta})\frac{\partial}{\partial\bm{\beta}}l(\bm{\xi}|\bm{\beta})^T+\frac{\partial^2}{\partial\bm{\beta}\partial\bm{\beta}^T}l(\bm{\xi}|\bm{\beta})\right)\right]_{|\bm{\beta}=\bm{0}}\cdot\bm{\beta}\nonumber\\
&=L(\bm{\xi}|\bm{\beta})_{|\bm{\beta}=\bm{0}}\left[1+\frac{\partial}{\partial\bm{\beta}}l(\bm{\xi}|\bm{\beta})^T_{|\bm{\beta}=\bm{0}}\cdot\bm{\beta}\nonumber\right.\\
&\left.\phantom{{}=L(\bm{\xi}|\bm{\beta})_{|\bm{\beta}=\bm{0}}\left[1\right.}+\frac{1}{2}\bm{\beta}^T\left[\frac{\partial}{\partial\bm{\beta}}l(\bm{\xi}|\bm{\beta})\frac{\partial}{\partial\bm{\beta}}l(\bm{\xi}|\bm{\beta})^T+\frac{\partial^2}{\partial\bm{\beta}\partial\bm{\beta}^T}l(\bm{\xi}|\bm{\beta})\right]_{|\bm{\beta}=\bm{0}}\cdot\bm{\beta}\right].\label{eq. method.4}
\end{align}
Plugging equation \eqref{eq. method.4} into \eqref{eq. method.13} yields
\begin{align*}
L(\eta,\bm{\xi})&=\int_{\bm{\beta}}L(\bm{\xi}|\bm{\beta})\bm H(\bm{\beta})\ d\bm{\beta}\\[0.8em]
&\approx L(\bm{\xi}|\bm{\beta})_{|\bm{\beta}=\bm{0}}\int_{\bm{\beta}}\left[1+\frac{\partial}{\partial\bm{\beta}}l(\bm{\xi}|\bm{\beta})^T_{|\bm{\beta}=\bm{0}}\cdot\bm{\beta}+\frac{1}{2}\bm{\beta}^T\left[\frac{\partial}{\partial\bm{\beta}}l(\bm{\xi}|\bm{\beta})\frac{\partial}{\partial\bm{\beta}}l(\bm{\xi}|\bm{\beta})^T+\frac{\partial^2}{\partial\bm{\beta}\partial\bm{\beta}^T}l(\bm{\xi}|\bm{\beta})\right]_{|\bm{\beta}=\bm{0}}\cdot\bm{\beta}\right]\bm H(\bm{\beta})\ d\bm{\beta} \\
&= L(\bm{\xi}|\bm{\beta})_{|\bm{\beta}=\bm{0}}\Big[1+\frac{\partial}{\partial\bm{\beta}}l(\bm{\xi}|\bm{\beta})^T_{|\bm{\beta}=\bm{0}}\cdot\text{E}[\bm{\beta}]
+\frac{1}{2}\text{tr}\Big\{\big[\frac{\partial}{\partial\bm{\beta}}l(\bm{\xi}|\bm{\beta})\frac{\partial}{\partial\bm{\beta}}l(\bm{\xi}|\bm{\beta})^T+\frac{\partial^2}{\partial\bm{\beta}\partial\bm{\beta}^T}l(\bm{\xi}|\bm{\beta})\big]_{|\bm{\beta}=\bm{0}}\cdot\text{E}[\bm{\beta}\bm{\beta}^T]\Big\}\Big].
\end{align*}
Since $\bm{\beta}$ is centered around $\bm{0_}{2r}$, we have
\begin{align}
\label{eq. method.5}
L(\eta,\bm{\xi})&\approx L(\bm{\xi}|\bm{\beta})_{|\bm{\beta}=\bm{0}}\Big[1+\frac{\partial}{\partial\bm{\beta}}l(\bm{\xi}|\bm{\beta})^T_{|\bm{\beta}=0}\cdot \bm{0}_{2r}+\frac{1}{2}\text{tr}\Big\{\big[\frac{\partial}{\partial\bm{\beta}}l(\bm{\xi}|\bm{\beta})\frac{\partial}{\partial\bm{\beta}}l(\bm{\xi}|\bm{\beta})^T+\frac{\partial^2}{\partial\bm{\beta}\partial\bm{\beta}^T}l(\bm{\xi}|\bm{\beta})\big]_{|\bm{\beta}=\bm{0}}\cdot\text{Var}[\bm{\beta}]\Big\}\Big] \nonumber \\
&= L(\bm{\xi}|\bm{\beta})_{|\bm{\beta}=\bm{0}}\left[1+\frac{1}{2}\text{tr}\left\{\left[\frac{\partial}{\partial\bm{\beta}}l(\bm{\xi}|\bm{\beta})\frac{\partial}{\partial\bm{\beta}}l(\bm{\xi}|\bm{\beta})^T+\frac{\partial^2}{\partial\bm{\beta}\partial\bm{\beta}^T}l(\bm{\xi}|\bm{\beta})\right]_{|\bm{\beta}=\bm{0}}\cdot\eta \Sigma_P \otimes \Sigma_G\right\}\right].
\end{align}
Using equation (\ref{eq. method.5}), we evaluate the score given by equation \eqref{eq:score},
\begin{align*}
U(\eta)
&\approx\frac{\partial}{\partial\eta}\text{log}\left(1+\frac{\eta}{2}\text{tr}\left\{\left[\frac{\partial}{\partial\bm{\beta}}l(\bm{\xi}|\bm{\beta})\frac{\partial}{\partial\bm{\beta}}l(\bm{\xi}|\bm{\beta})^T+\frac{\partial^2}{\partial\bm{\beta}\partial\bm{\beta}^T}l(\bm{\xi}|\bm{\beta})\right]_{|\bm{\beta}=\bm{0}}\cdot \Sigma_P \otimes \Sigma_G\right\}\right).\nonumber
\end{align*}
Using Taylor's expansion of $\text{log}(1+x)$ around $x=0$ yields
\begin{align}
U(\eta)&\approx\frac{\partial}{\partial\eta}\left[\frac{\eta}{2}\text{tr}\left\{\left[\frac{\partial}{\partial\bm{\beta}}l(\bm{\xi}|\bm{\beta})\frac{\partial}{\partial\bm{\beta}}l(\bm{\xi}|\bm{\beta})^T+\frac{\partial^2}{\partial\bm{\beta}\partial\bm{\beta}^T}l(\bm{\xi}|\bm{\beta})\right]_{|\bm{\beta}=0}\cdot\Sigma_P \otimes \Sigma_G\right\}\right]\nonumber\\
&=\frac{1}{2}\text{tr}\left\{\frac{\partial}{\partial\bm{\beta}}l(\bm{\xi}|\bm{\beta})\cdot\frac{\partial}{\partial\bm{\beta}}l(\bm{\xi}|\bm{\beta})^T|_{\bm{\beta}=\bm{0}}\cdot\Sigma_P \otimes \Sigma_G+\frac{\partial^2}{\partial\bm{\beta}\partial\bm{\beta}^T}l(\bm{\xi}|\bm{\beta})|_{\bm{\beta}=\bm{0}}\cdot\Sigma_P \otimes \Sigma_G\right\}\nonumber \\
&=\frac{1}{2}\frac{\partial }{\partial\bm{\beta}}l(\bm{\xi}|\bm{\beta})^T|_{\bm{\beta}=\bm{0}}\left(\Sigma_P \otimes \Sigma_G\right)\frac{\partial}{\partial\bm{\beta}}l(\bm{\xi}|\bm{\beta})|_{\bm{\beta}=\bm{0}}+\frac{1}{2}\text{tr}\left\{\frac{\partial^2}{\partial\bm{\beta}\bm{\beta}^T}l(\bm{\xi}|\bm{\beta})|_{\bm{\beta}=\bm{0}}\cdot\Sigma_P \otimes \Sigma_G\right\}.\label{eq. method.7}
\end{align}
Using the more compact notations derived in \eqref{eq:compact1} and \eqref{eq:compact2} for the gradient and Hessian of the conditional log-likelihood respectively, the score test for $H_0:\eta=0$ can finally be rewritten as
\begin{align*}
U(\eta)
&= \frac{1}{2}\bm{L}^T(I_2\otimes \bm{G})\left(\Sigma_P\otimes\Sigma_G\right)(I_2\otimes \bm{G}^T)\bm{L}+\frac{1}{2}\text{tr}\left\{ \left(I_2\otimes \bm{G}^T\right) D \left(I_2\otimes \bm{G}\right) (\Sigma_P\otimes\Sigma_G) \right\} \\
&= \frac{1}{2}\Big(\bm{L}^T(\Sigma_P \otimes \bm{G}\Sigma_G \bm{G}^T)\bm{L} + \text{tr}\left\{ (\Sigma_P \otimes \bm{G}\Sigma_G \bm{G}^T) \bm D\right\}\Big).
\end{align*}

\clearpage
\section*{Appendix B. Variance-Covariance Matrix Correction for plug-in estimates of nuisance parameters}

Let $l(\bm{\xi}|\bm{\beta})$ be the log of the conditional likelihood given $\bm{\beta}$, which depends on $\bm{\beta}\in\mathbbm{R}^{2r}$ and a vector of nuisance parameters $\bm{\xi}=(\theta,\bm{\gamma_}1,\bm{\gamma_}2,\phi_1,\phi_2)^T\in\mathbbm{R}^{2m+3}$, where $\theta\in\mathbbm{R}$ is the copula parameter, while  $\bm{\gamma_}1, \bm{\gamma_}2 \in \mathbbm{R}^m$ and $\phi_1,\phi_2 \in\mathbbm{R}$ are respectively the covariates fixed effects and dispersion parameters of the marginal models. Under  the  null  model $H_0:\bm{\beta}=0_{2r}$, assuming the vector of nuisance parameters $\bm \xi$ is known, it follows from standard asymptotic theory that the score ${\frac{\partial }{\partial\bm{\beta}}l(\bm{\xi}|\bm{\beta})=(I_2\otimes \bm{G}^T)L}$ follows an $2r$-variate normal distribution with zero mean and variance-covariance matrix given by the expected information matrix $$-\mathbbm{E}_{\bm{\beta}}[\frac{\partial}{\partial\bm{\beta}\partial\bm{\beta}^T}l(\bm{\xi}|\bm{\beta})]=-\mathbbm{E}_{\bm{\beta}}[(I_2\otimes\bm{G}^T)\bm{D}(I_2\otimes\bm{G})].$$ We can replace the expected information matrix by the observed information matrix $$I_{\bm{\beta\beta}}=-\frac{\partial}{\partial\bm{\beta}\partial\bm{\beta}^T}l(\bm{\xi}|\bm{\beta})=-(I_2\otimes\bm{G}^T)\bm{D}(I_2\otimes\bm{G})$$ since it can be shown to be a consistent estimator. Thus, it is easy to show that the asymptotic distribution of $U(\eta)=\frac{1}{2}\Big(\bm{L}^T(\Sigma_P \otimes \bm{G}\Sigma_G \bm{G}^T)\bm{L} + \text{tr}\left\{ (\Sigma_P \otimes \bm{G}\Sigma_G \bm{G}^T) \bm D\right\}\Big)$ is equivalent to that of $\sum_{i=1}^{2r}\lambda_i(\chi^2_1 - 1)/2$, where $\lambda_i$, $i=1,...,2r$, are eigenvalues of ${I_{\bm{\beta\beta}}^{1/2}(\Sigma_P \otimes \Sigma_G)I_{\bm{\beta\beta}}^{1/2}}$.

However, when we estimate the nuisance parameters in $\xi$ under $H_0$ and plug their estimates into the expected information matrix $I_{\bm{\beta\beta}}$, it can lead to severe type I error inflation of the proposed score test. Indeed, one needs an appropriate correction for the variability induced by the plug-in estimates $\hat\xi$ in the asymptotic variance-covariance of the score $(I_2\otimes \bm{G}^T)L$. The score vector for all parameters is given by 
\begin{equation*}
\bm{U}(\bm{\bm{\beta}},\bm{\bm{\xi}})=\begin{bmatrix} \frac{\partial{l(\bm{\xi}|\bm{\beta})}}{\partial\bm{\beta}}^T & \frac{\partial{l(\bm{\xi}|\bm{\beta}))}}{\partial\bm{\xi}}^T \end{bmatrix}^T.
\end{equation*}
The observed Fisher information matrix is 
\begin{align*}
\bm{I}_{\text{obs}}(\bm{\beta},\bm{\xi})&=-\frac{\partial^2l(\bm{\xi}|\bm{\beta})}{\partial\bm{\psi}\partial\bm{\psi}^T} \nonumber\\
&=\begin{bmatrix}\bm{I}_{\bm{\beta}\bm{\beta}}(\bm{\beta},\bm{\xi})&\bm{I}_{\bm{\beta}\bm{\xi}}(\bm{\beta},\bm{\xi})\\\bm{I}_{\bm{\xi}\bm{\beta}}(\bm{\beta},\bm{\xi})&\bm{I}_{\bm{\xi}\bm{\xi}}(\bm{\beta},\bm{\xi}) \end{bmatrix} ,
\end{align*} 
with $\bm{\psi}=(\bm{\beta}^T,\bm{\xi}^T)^T$ and noting its inverse
\begin{equation*}
\bm{I}_{\text{obs}}(\bm{\beta},\bm{\xi})^{-1}=\begin{bmatrix}\bm{I}^{\bm{\beta}\bm{\beta}}(\bm{\beta},\bm{\xi})&\bm{I}^{\bm{\beta}\bm{\xi}}(\bm{\beta},\bm{\xi})\\\bm{I}^{\bm{\xi}\bm{\beta}}(\bm{\beta},\bm{\xi})&\bm{I}^{\bm{\xi}\bm{\xi}}(\bm{\beta},\bm{\xi}) \end{bmatrix}.
\end{equation*}
It follows that (\citealp{inverse})
\begin{equation}
\label{inversefisher}
\bm{I}^{\bm{\beta}\bm{\beta}}(\bm{\beta},\bm{\xi})=\bm{I}_{\bm{\beta}\bm{\beta}}(\bm{\beta},\bm{\xi})^{-1}+\bm{I}_{\bm{\beta}\bm{\beta}}(\bm{\beta},\bm{\xi})^{-1}\bm{I}_{\bm{\beta}\bm{\xi}}(\bm{\beta},\bm{\xi})\bm{Z}^{-1}\bm{I}_{\bm{\xi}\bm{\beta}}(\bm{\beta},\bm{\xi})\bm{I}_{\bm{\beta}\bm{\beta}}(\bm{\beta},\bm{\xi})^{-1},
\end{equation}
with $$\bm{Z}=\bm{I}_{\bm{\xi}\bm{\xi}}(\bm{\beta},\bm{\xi})-\bm{I}_{\bm{\xi}\bm{\beta}}(\bm{\beta},\bm{\xi})\bm{I}_{\bm{\beta}\bm{\beta}}(\bm{\beta},\bm{\xi})^{-1}\bm{I}_{\bm{\beta}\bm{\xi}}(\bm{\beta},\bm{\xi})$$
and $$\bm{I}_{\bm{\beta}\bm{\xi}}(\bm{\beta},\bm{\xi})=\bm{I}_{\bm{\xi}\bm{\beta}}(\bm{\beta},\bm{\xi})=-\frac{\partial^2l(\bm{\xi}|\bm{\beta})}{\partial\bm{\xi}\partial\bm{\beta}^T}.$$
The statistic for the score test under $H_0:\bm{\beta}={\bm{\beta_}0}$ is given by 
\begin{equation*}
\frac{\partial l}{\partial\bm{\beta}}(\hat{\bm{\xi}}|\bm{\beta_}0)^T\bm{I}^{\bm{\beta}\bm{\beta}}({\bm{\beta_}0}_,\hat{\bm{\xi}})\frac{\partial l}{\partial\bm{\beta}}(\hat{\bm{\xi}}|\bm{\beta_}0)\underset{H_0}{\sim} \chi^2_{2r} ,
\end{equation*}
with $\hat{\bm{\xi}}$ estimated under $H_0$.
In other words, we have
\begin{equation*}
\frac{\partial l}{\partial\bm{\beta}}(\hat{\bm{\xi}}|\bm{\beta_}0)\underset{H_0}{\sim} N_{2r}(\bm{0_}{2r},[\bm{I}^{\bm{\beta}\bm{\beta}}({\bm{\beta_}0},\hat{\bm{\xi}})]^{-1})\label{eq. fisher.1}.
\end{equation*}
Thus, $[\bm{I}^{\bm{\beta}\bm{\beta}}({\bm{\beta_}0},\hat{\bm{\xi}})]^{-1}$ defined in \eqref{inversefisher} is a more consistant variance-covariance matrix for the score $(I_2\otimes \bm{G}^T)L$, correcting for the plug-in of the estimates $\hat\xi$ in the score calculation.

\clearpage
\section*{Appendix C. Heritability}
Heritability of a trait is defined as the fraction of the total phenotypic variability that is attributable to genetic variability. For our mixed generalized linear model, the heritability for the $ k ^ {th} $ trait is given by
\begin{align}
h^2_k&=\frac{\text{Var}[\bm{G_}i^T\bm{\beta_}k]}{\text{Var}[Y_{ik}]}\nonumber\\
&=\frac{\text{E}[\eta\ \bm{G_}i^T\bm{W}\bm{G_}i]}{\text{Var}[g_k^{-1}\left(\bm{X_}i^T\bm{\gamma}_j+\bm{G}_i^T\bm{\beta_}k\right)]+\text{E}[\phi_k\nu(\mu_{ik})]}\label{eq. Puissance.1}
\end{align}
since $\bm{\beta_k}\sim F(\bm{0_}r,\eta\bm{W})$. In general, we cannot find any closed form for equation \eqref{eq. Puissance.1}, except for linear models where it simplifies to
\begin{align}
h^2_k&=\frac{\eta\cdot\text{E}\left[\sum_{j=1}^{r}w_jg_{ij}^2\right]}{\text{Var}[\bm{X_}i^T\bm{\gamma_}k+\bm{G}_i^T\bm{\beta_}k]+\phi_k}\nonumber\\
&=\frac{\eta\cdot\sum_{j=1}^{r}w_j\text{E}[g_{ij}^2]}{\bm{\gamma_}k^T\text{Var}[\bm{X_}i]\bm{\gamma}_k + \eta\cdot\sum_{j=1}^{r}w_j\text{E}[g_{ij}^2] + \phi_k}\nonumber\\
&=\frac{2\eta\cdot\sum_{j=1}^{r}w_jp_j(p_j+1)}{2\eta\cdot\sum_{j=1}^{r}w_jp_j(p_j+1)+\phi_k^*}\label{eq. Puissance.2}
\end{align}
where $\phi_k^* = \bm{\gamma_}k^T\text{Var}[\bm{X_}i]\bm{\gamma}_k + \phi_k$, and we assumed $g_{ij}\sim\textit{Binomial}(2,p_j)$ with $p_j$ the minor allele frequency for $j^{th}$ locus, $j=1,..,r$.

We isolate $\eta$ in \eqref{eq. Puissance.2}, from where
\begin{align}
\eta=\frac{\phi_k^* h_k^2}{2(1-h_k^2)\sum_{j=1}^{r}w_jp_j(p_j+1)}, \label{eq. Puissance.4}
\end{align}
where we set $w_j=1$ for randomly selected causal variants, and $w_j=0$ otherwise. Thus, $\eta$ depends on the heritability $h_k^2$ and the fraction of causal variants.

\clearpage
\section*{Appendix D. Power of CBMAT with a resampling based approach for deriving p-values}

Consider the series of statistics $Q_{\rho_j}$ defined in Section 2.3 as
\begin{align*}
\nonumber
Q_{\rho_j} &= \bm{L}^T (\Sigma_{\rho_j} \otimes \bm{G} \Sigma_G \bm{G}^T) \bm{L} \\
           & = \bm{Z}^T \tilde{\bm{K}}_{\rho_j} \bm{Z},
\end{align*}
where $\bm{Z} = \tilde{\bm{B}}^{-1/2} \left(I_2\otimes \bm{G}^T\right) \bm{L} \sim N(\bm 0, \bm I_{2r})$, and $\tilde{\bm{K}}_{\rho_j} = \tilde{\bm{B}}^{1/2}(\Sigma_{\rho_j} \otimes \Sigma_G)\tilde{\bm{B}}^{1/2}$ for $j=1,\ldots, b$. We evaluated the power of CBMAT when deriving p-values via a resampling based approach to estimate the correlation structure of the $Q_{\rho_j}$ statistics. That is, we first generated $2N$ samples from an $N(0, 1)$ distribution, say $Z_R$, and then calculated $b$ different test statistics as $Q_{R{_j}}=Z_R^T\tilde{\bm{B}}^{1/2}(\Sigma_{\rho_j} \otimes \Sigma_G)\tilde{\bm{B}}^{1/2}Z_R$ for $j=1,\ldots, b$. We repeated the previous steps independently for $R=1000$ iterations, and calculated the Kendall's tau correlation matrix $\Gamma$ between $(Q_{R_{1}}, Q_{R_{2}},...,Q_{R_{b}})$ from the $R$ resampled statistics. We then used a Gaussian copula with the estimated null correlation structure to approximate the joint null distribution of observed statistics $(Q_{\rho_{1}}, Q_{\rho_{2}},...,Q_{\rho_{b}})$ and derived a resampling based approach p-value for our score statistics. \\

\begin{figure}[!p]
\centering
\includegraphics[width=1\textwidth]{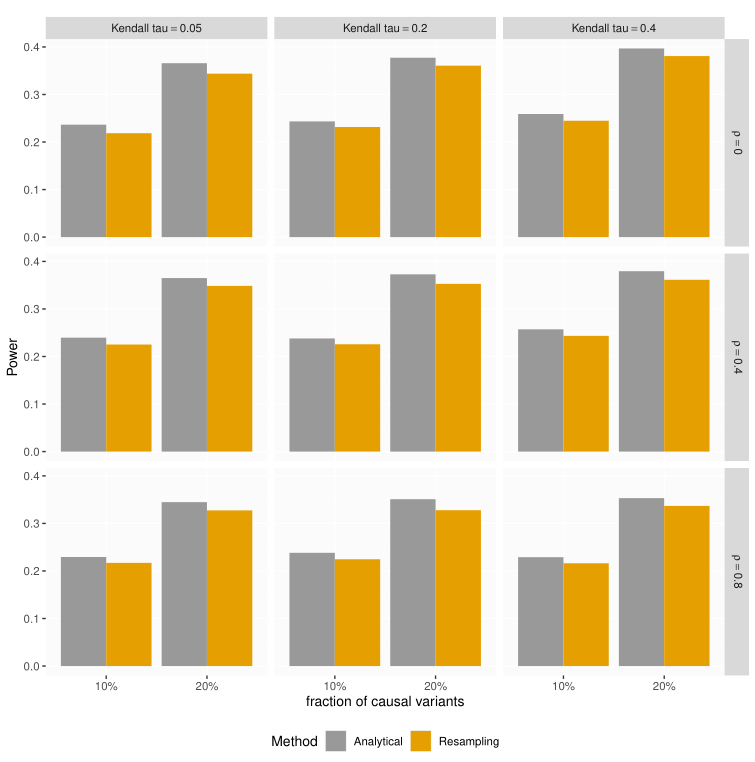}
\caption{Power of CBMAT to detect SNPs/phenotype association using analytical and resampling based approach. Results are computed from 5 000 data sets generated under a Gaussian copula model when either $10\%$ or $20\%$ of the variants are causal and when both traits are continuous and follow Exponential/Student-t marginal distributions. The variance component parameter $\eta$ was fixed such that the traits heritabilty under a linear model is approximately $2\%$.}

\label{fig:power.5}
\end{figure}

\begin{figure}[!p]
\centering
\includegraphics[width=1\textwidth]{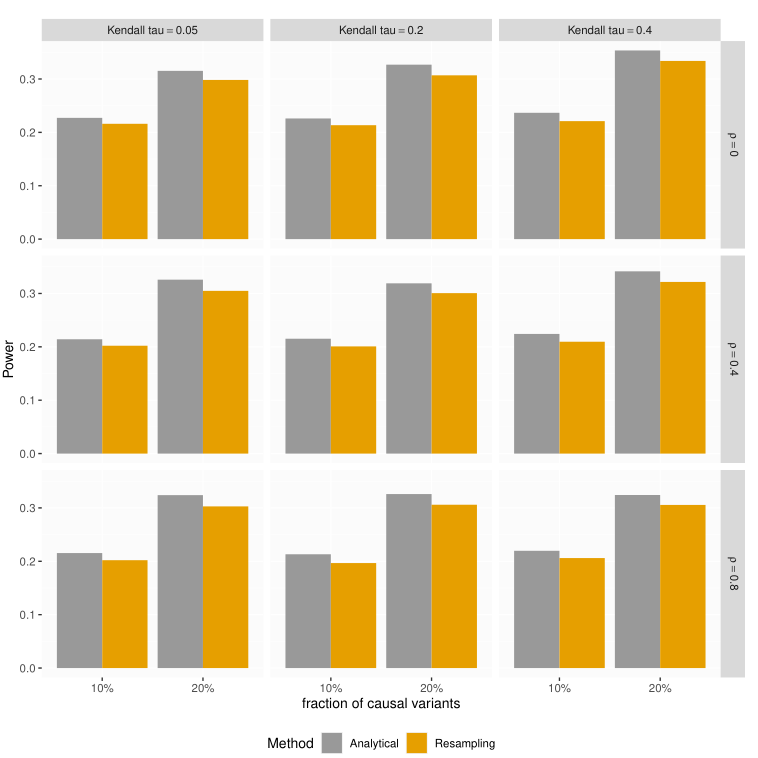}
\caption{Power of CBMAT to detect SNPs/phenotype association using analytical and resampling based approach. Results are computed from 5 000 data sets generated under a Gaussian copula model when either $10\%$ or $20\%$ of the variants are causal and when traits are mixed binary-continuous and follow Probit/Exponential marginal distributions. The variance component parameter $\eta$ was fixed such that the traits heritabilty under a linear model is approximately $2\%$.}
\label{fig:power.6}
\end{figure}

To evaluate the power of CBMAT, we considered again genetic variants located within 500kbs of the BRCA1 gene, using a total of $N=503$ subjects with European ancestry from the 1000 Genomes Project. A genetic region composed of $ r = 30 $ consecutive common/rare variants, randomly selected at each iteration, with equal weights $ w_j = 1$ was considered. We set the number of replications to 5000 and used the Gaussian copula to simulate either a bivariate continuous phenotype, with Exponential and Student-t correlated marginal distributions, or a mixed binary continuous phenotype, with binary and Student-t correlated marginal distributions. We compared different scenarios where the pleiotropic correlation parameter $\rho$ was equal to $\{0, 0.4, 0.8\}$ and the Kendall's $\tau$ was equal to $\{0.05,0.2,0.4\}$. Furthermore, we were interested in situations for which respectively $10\% $ and $ 20\% $ of the variants were causal. Results for continuous and mixed bivariate phenotypes are presented in Figure \ref{fig:power.5} and Figure \ref{fig:power.6} respectively. As can be seen from these results, our procedure  for obtaining  an  analytic  p-value  of  the  proposed score test yields slightly greater power than the resampling based approach for all simulations scenarios. Thus, not only does the use of Pearson correlation matrix to model dependence between the test statistics effectively controls type I error, but it also results in a slight uniform increase of power to detect association.

\clearpage
\section*{Appendix E. Supplementary Tables}


\begin{table}[!h]
\centering
\caption{Power $(\%)$ to detect SNPs/phenotype association where the data are generated under either Gaussian or Clayton copula models. Results are computed from 5 000 data sets generated when either $10\%$ or $20\%$ of the variants are causal and both traits are continuous and both follow Exponential marginal distributions. The variance component parameter $\eta$ was fixed such that the traits heritabilty under a linear model is approximately $2\%$. The compared methods are CBMAT fitted with true marginal models (True), and fitted using marginal models chosen based on AIC (AIC);  MURAT and Multi-SKAT are fitted without INT transformation (no INT) and with INT transformation after adjusting for the covariates (INT); Multi-SKAT is fitted with homogeneous (Multi-SKAT Hom) and uncorrelated (Multi-SKAT Het) effect sizes.}
\label{tab:power.3}

\resizebox{\textwidth}{!}{
\begin{tabular}{cccccccccccc}
  \hline & & & & \multicolumn{2}{c}{CBMAT} & \multicolumn{2}{c}{MURAT}  & \multicolumn{2}{c}{Multi-SKAT Het} & \multicolumn{2}{c}{Multi-SKAT Hom} \\
 copula & $\rho$ & Kendall's $\tau$ & \% causal & True & AIC & no INT & INT & no INT & INT & no INT & INT \\
  \hline
Normal & 0.0 & 0.05 & 10\% & 15.3 & 15.3 & 8.1 & 5.7 & 19.0 & 8.5 & 5.9 & 4.5 \\ 
   &  &  & 20\% & 17.5 & 17.5 & 11.2 & 8.8 & 17.7 & 9.9 & 5.5 & 4.4 \\ 
   &  & 0.20 & 10\% & 15.4 & 15.4 & 8.2 & 6.7 & 22.1 & 9.8 & 6.7 & 3.5 \\ 
   &  &  & 20\% & 18.5 & 18.5 & 11.7 & 10.4 & 19.8 & 12.1 & 5.1 & 4.2 \\ 
   &  & 0.40 & 10\% & 19.8 & 19.9 & 10.8 & 8.3 & 26.9 & 14.8 & 5.9 & 3.1 \\ 
   &  &  & 20\% & 22.7 & 22.7 & 14.0 & 12.6 & 25.6 & 17.7 & 5.1 & 3.5 \\ 
   \hline
 & 0.4 & 0.05 & 10\% & 14.6 & 14.6 & 7.8 & 5.5 & 16.6 & 6.8 & 7.3 & 5.5 \\ 
   &  &  & 20\% & 16.6 & 16.6 & 10.2 & 7.6 & 15.6 & 7.9 & 6.5 & 6.7 \\ 
   &  & 0.20 & 10\% & 14.7 & 14.5 & 8.4 & 5.9 & 18.9 & 7.1 & 7.4 & 5.0 \\ 
   &  &  & 20\% & 16.1 & 16.1 & 10.5 & 7.9 & 17.7 & 9.1 & 6.7 & 5.5 \\ 
   &  & 0.40 & 10\% & 16.2 & 16.2 & 9.4 & 7.0 & 23.0 & 11.0 & 7.4 & 4.8 \\ 
   &  &  & 20\% & 18.8 & 18.8 & 12.2 & 10.5 & 20.5 & 11.9 & 6.3 & 4.7 \\ 
   \hline
 & 0.8 & 0.05 & 10\% & 14.3 & 14.3 & 7.5 & 4.8 & 14.5 & 5.0 & 9.3 & 7.1 \\ 
   &  &  & 20\% & 16.5 & 16.5 & 9.9 & 7.3 & 12.8 & 5.8 & 8.0 & 8.5 \\ 
   &  & 0.20 & 10\% & 13.1 & 13.1 & 7.1 & 4.7 & 15.0 & 4.3 & 9.6 & 6.8 \\ 
   &  &  & 20\% & 14.9 & 14.9 & 9.2 & 6.9 & 13.5 & 5.2 & 7.8 & 7.3 \\ 
   &  & 0.40 & 10\% & 12.7 & 12.7 & 6.9 & 5.0 & 16.4 & 5.6 & 8.9 & 5.9 \\ 
   &  &  & 20\% & 14.4 & 14.4 & 8.8 & 7.0 & 14.6 & 5.7 & 7.7 & 6.1 \\ 
   \hline
Clayton & 0.0 & 0.05 & 10\% & 15.4 & 15.4 & 8.1 & 5.7 & 19.3 & 8.4 & 6.5 & 4.6 \\ 
   &  &  & 20\% & 17.6 & 17.6 & 11.2 & 8.7 & 18.2 & 9.9 & 5.8 & 4.5 \\ 
   &  & 0.20 & 10\% & 15.2 & 15.2 & 7.9 & 6.5 & 20.3 & 9.7 & 5.9 & 3.8 \\ 
   &  &  & 20\% & 17.7 & 17.7 & 11.0 & 10.3 & 19.4 & 12.0 & 5.1 & 4.2 \\ 
   &  & 0.40 & 10\% & 17.8 & 18.0 & 9.3 & 8.0 & 24.1 & 13.3 & 5.9 & 3.2 \\ 
   &  &  & 20\% & 21.7 & 21.7 & 11.6 & 11.7 & 22.0 & 14.9 & 5.1 & 3.4 \\ 
   \hline
 & 0.4 & 0.05 & 10\% & 15.0 & 15.0 & 7.9 & 5.5 & 16.3 & 6.2 & 7.2 & 5.6 \\ 
   &  &  & 20\% & 16.8 & 16.8 & 10.3 & 8.0 & 15.8 & 7.9 & 6.7 & 6.9 \\ 
   &  & 0.20 & 10\% & 16.0 & 15.9 & 8.4 & 5.9 & 17.8 & 7.0 & 7.3 & 5.4 \\ 
   &  &  & 20\% & 16.2 & 16.2 & 9.9 & 7.6 & 17.6 & 8.6 & 6.4 & 5.8 \\ 
   &  & 0.40 & 10\% & 16.8 & 16.8 & 8.0 & 6.0 & 20.2 & 9.0 & 7.2 & 4.4 \\ 
   &  &  & 20\% & 18.3 & 18.3 & 10.6 & 9.2 & 17.8 & 11.3 & 5.8 & 5.0 \\ 
   \hline
 & 0.8 & 0.05 & 10\% & 14.3 & 14.3 & 7.5 & 4.7 & 14.8 & 5.1 & 9.9 & 7.7 \\ 
   &  &  & 20\% & 16.3 & 16.3 & 9.7 & 7.2 & 12.6 & 5.8 & 7.7 & 8.6 \\ 
   &  & 0.20 & 10\% & 14.9 & 14.9 & 7.8 & 4.9 & 15.6 & 4.8 & 9.9 & 7.1 \\ 
   &  &  & 20\% & 16.0 & 16.0 & 8.9 & 7.1 & 13.5 & 4.8 & 7.6 & 7.8 \\ 
   &  & 0.40 & 10\% & 14.1 & 14.1 & 6.9 & 4.4 & 15.9 & 4.6 & 8.7 & 5.6 \\ 
   &  &  & 20\% & 16.2 & 16.2 & 9.0 & 6.7 & 14.4 & 5.0 & 7.5 & 6.4 \\ 
   \hline
\end{tabular}}
\end{table}

\begin{table}[!h]
\centering
\caption{Power $(\%)$ to detect SNPs/phenotype association where the data are generated under either Gaussian or Clayton copula models. Results are computed from 5 000 data sets generated when either $10\%$ or $20\%$ of the variants are causal and both traits are continuous and follow Exponential/Student-t marginal distributions. The variance component parameter $\eta$ was fixed such that the traits heritabilty under a linear model is approximately $2\%$. The compared methods are CBMAT fitted with true marginal models (True), and fitted using marginal models chosen based on AIC (AIC);  MURAT and Multi-SKAT are fitted without INT transformation (no INT) and with INT transformation after adjusting for the covariates (INT); Multi-SKAT is fitted with homogeneous (Multi-SKAT Hom) and uncorrelated (Multi-SKAT Het) effect sizes.}
\label{tab:power.4}

\resizebox{\textwidth}{!}{
\begin{tabular}{cccccccccccc}
  \hline & & & & \multicolumn{2}{c}{CBMAT} & \multicolumn{2}{c}{MURAT}  & \multicolumn{2}{c}{Multi-SKAT Het} & \multicolumn{2}{c}{Multi-SKAT Hom} \\
 copula & $\rho$ & Kendall's $\tau$ & \% causal & True & AIC & no INT & INT & no INT & INT & no INT & INT \\
  \hline
Normal & 0.0 & 0.05 & 10\% & 13.0 & 13.0 & 8.9 & 9.3 & 9.5 & 13.8 & 2.1 & 8.2 \\ 
   &  &  & 20\% & 18.8 & 18.8 & 12.2 & 14.3 & 9.6 & 15.6 & 2.0 & 9.6 \\ 
   &  & 0.20 & 10\% & 14.4 & 14.4 & 9.2 & 9.7 & 11.3 & 13.8 & 2.3 & 6.3 \\ 
   &  &  & 20\% & 21.0 & 21.0 & 13.3 & 15.2 & 12.1 & 17.5 & 2.2 & 7.9 \\ 
   &  & 0.40 & 10\% & 17.3 & 17.3 & 9.6 & 11.0 & 15.4 & 16.8 & 2.8 & 5.4 \\ 
   &  &  & 20\% & 26.1 & 26.1 & 14.2 & 17.8 & 14.9 & 20.3 & 2.7 & 6.2 \\ 
   \hline
 & 0.4 & 0.05 & 10\% & 13.4 & 13.4 & 9.2 & 9.6 & 7.3 & 13.1 & 2.6 & 11.1 \\ 
   &  &  & 20\% & 19.8 & 19.8 & 12.9 & 15.0 & 7.2 & 14.9 & 2.7 & 12.6 \\ 
   &  & 0.20 & 10\% & 13.5 & 13.5 & 9.1 & 9.5 & 8.6 & 11.2 & 2.5 & 9.6 \\ 
   &  &  & 20\% & 18.6 & 18.6 & 12.3 & 14.0 & 8.5 & 14.0 & 2.7 & 11.3 \\ 
   &  & 0.40 & 10\% & 15.6 & 15.6 & 9.0 & 10.1 & 11.1 & 12.4 & 3.1 & 7.6 \\ 
   &  &  & 20\% & 22.3 & 22.3 & 12.5 & 15.3 & 11.0 & 14.0 & 2.6 & 9.4 \\ 
   \hline
 & 0.8 & 0.05 & 10\% & 13.5 & 13.5 & 9.0 & 9.6 & 5.2 & 11.3 & 3.4 & 13.3 \\ 
   &  &  & 20\% & 18.8 & 18.8 & 13.2 & 14.1 & 5.9 & 14.5 & 3.6 & 16.9 \\ 
   &  & 0.20 & 10\% & 12.9 & 12.8 & 8.4 & 8.5 & 6.1 & 9.2 & 3.3 & 12.4 \\ 
   &  &  & 20\% & 17.4 & 17.4 & 12.0 & 13.1 & 5.9 & 10.6 & 2.8 & 13.9 \\ 
   &  & 0.40 & 10\% & 12.4 & 12.4 & 8.0 & 8.4 & 7.8 & 7.4 & 3.3 & 10.5 \\ 
   &  &  & 20\% & 18.8 & 18.8 & 11.1 & 12.3 & 7.5 & 8.3 & 2.9 & 12.0 \\ 
   \hline
Clayton & 0.0 & 0.05 & 10\% & 12.6 & 12.6 & 8.8 & 9.1 & 9.1 & 13.5 & 2.1 & 8.2 \\ 
   &  &  & 20\% & 18.8 & 18.8 & 13.0 & 14.3 & 9.2 & 15.6 & 1.6 & 9.6 \\ 
   &  & 0.20 & 10\% & 14.4 & 14.4 & 9.3 & 9.8 & 10.8 & 13.9 & 2.0 & 7.0 \\ 
   &  &  & 20\% & 20.6 & 20.6 & 12.5 & 14.8 & 10.0 & 16.1 & 2.2 & 8.1 \\ 
   &  & 0.40 & 10\% & 16.9 & 16.9 & 9.3 & 10.3 & 14.1 & 16.2 & 2.9 & 5.8 \\ 
   &  &  & 20\% & 26.4 & 26.4 & 14.2 & 17.8 & 14.0 & 20.2 & 2.1 & 7.5 \\ 
   \hline
 & 0.4 & 0.05 & 10\% & 13.6 & 13.6 & 8.7 & 9.6 & 7.3 & 13.2 & 2.8 & 11.5 \\ 
   &  &  & 20\% & 18.7 & 18.7 & 12.5 & 14.1 & 6.8 & 14.1 & 2.4 & 12.5 \\ 
   &  & 0.20 & 10\% & 13.2 & 13.2 & 8.8 & 9.0 & 8.4 & 12.3 & 2.6 & 9.7 \\ 
   &  &  & 20\% & 19.3 & 19.3 & 12.7 & 14.3 & 8.3 & 13.7 & 2.5 & 11.5 \\ 
   &  & 0.40 & 10\% & 15.8 & 15.8 & 9.0 & 9.8 & 10.4 & 12.3 & 2.7 & 8.1 \\ 
   &  &  & 20\% & 22.7 & 22.7 & 12.3 & 14.5 & 10.2 & 13.7 & 2.7 & 9.9 \\ 
   \hline
 & 0.8 & 0.05 & 10\% & 13.5 & 13.5 & 9.3 & 9.9 & 5.6 & 11.2 & 3.5 & 13.8 \\ 
   &  &  & 20\% & 18.9 & 18.9 & 13.0 & 14.4 & 5.5 & 14.5 & 3.5 & 17.3 \\ 
   &  & 0.20 & 10\% & 13.0 & 13.0 & 9.0 & 8.5 & 5.8 & 9.0 & 3.0 & 12.1 \\ 
   &  &  & 20\% & 17.6 & 17.6 & 12.2 & 12.9 & 5.4 & 11.0 & 3.1 & 14.7 \\ 
   &  & 0.40 & 10\% & 13.5 & 13.5 & 8.5 & 8.5 & 7.6 & 8.0 & 3.3 & 10.8 \\ 
   &  &  & 20\% & 19.7 & 19.7 & 12.0 & 12.6 & 6.9 & 8.5 & 3.1 & 11.7 \\ 
   \hline
\end{tabular}}
\end{table}

\begin{table}[!h]
\centering
\caption{Power $(\%)$ to detect SNPs/phenotype association where the data are generated under either Gaussian or Clayton copula models. Results are computed from 5 000 data sets generated when either $10\%$ or $20\%$ of the variants are causal and traits are mixed binary-continuous and follow Probit/Exponential marginal distributions. The variance component parameter $\eta$ was fixed such that the traits heritabilty under a linear model is approximately $2\%$. The compared methods are CBMAT fitted with true marginal models (True), and fitted using marginal models chosen based on AIC (AIC);  MURAT and Multi-SKAT are fitted without INT transformation (no INT) and with INT transformation after adjusting for the covariates (INT); Multi-SKAT is fitted with homogeneous (Multi-SKAT Hom) and uncorrelated (Multi-SKAT Het) effect sizes.}
\label{tab:power.1}

\resizebox{\textwidth}{!}{
\begin{tabular}{llllcccccccc}
  \hline & & & & \multicolumn{2}{c}{CBMAT} & \multicolumn{2}{c}{MURAT}  & \multicolumn{2}{c}{Multi-SKAT Het} & \multicolumn{2}{c}{Multi-SKAT Hom} \\
 copula & $\rho$ & Kendall's $\tau$ & \% causal & True & AIC & no INT & INT & no INT & INT & no INT & INT \\
  \hline
Normal & 0.0 & 0.05 & 10\% & 10.2 & 10.2 & 6.4 & 5.0 & 1.4 & 5.9 & 0.5 & 3.9 \\ 
   &  &  & 20\% & 10.7 & 10.7 & 7.3 & 6.7 & 1.6 & 7.1 & 0.6 & 4.3 \\ 
   &  & 0.20 & 10\% & 11.1 & 11.1 & 6.3 & 5.1 & 2.1 & 6.7 & 0.8 & 4.2 \\ 
   &  &  & 20\% & 12.5 & 12.5 & 8.2 & 7.4 & 2.6 & 7.9 & 0.9 & 3.7 \\ 
   &  & 0.40 & 10\% & 12.0 & 12.0 & 6.5 & 5.7 & 3.3 & 7.5 & 1.0 & 3.1 \\ 
   &  &  & 20\% & 14.1 & 14.1 & 8.4 & 7.9 & 3.5 & 8.6 & 1.4 & 3.0 \\ 
   \hline
 & 0.4 & 0.05 & 10\% & 10.4 & 10.4 & 7.0 & 5.3 & 0.9 & 6.1 & 0.5 & 5.7 \\ 
   &  &  & 20\% & 10.8 & 10.8 & 7.1 & 6.5 & 1.0 & 6.3 & 0.5 & 6.4 \\ 
   &  & 0.20 & 10\% & 10.2 & 10.2 & 6.2 & 4.8 & 1.4 & 5.2 & 0.6 & 4.5 \\ 
   &  &  & 20\% & 11.4 & 11.4 & 7.6 & 6.7 & 1.2 & 6.1 & 0.5 & 4.8 \\ 
   &  & 0.40 & 10\% & 10.6 & 10.6 & 5.8 & 4.3 & 2.0 & 5.3 & 0.6 & 4.4 \\ 
   &  &  & 20\% & 12.3 & 12.3 & 7.7 & 6.4 & 2.0 & 5.8 & 0.7 & 4.4 \\ 
   \hline
 & 0.8 & 0.05 & 10\% & 11.1 & 11.1 & 7.0 & 5.5 & 0.4 & 4.9 & 0.4 & 6.9 \\ 
   &  &  & 20\% & 12.1 & 12.1 & 8.4 & 7.1 & 0.5 & 6.1 & 0.4 & 8.2 \\ 
   &  & 0.20 & 10\% & 10.0 & 10.0 & 6.3 & 4.8 & 0.6 & 3.9 & 0.4 & 5.9 \\ 
   &  &  & 20\% & 11.0 & 11.0 & 7.9 & 6.4 & 0.7 & 4.4 & 0.4 & 6.8 \\ 
   &  & 0.40 & 10\% & 9.8 & 9.8 & 5.8 & 4.3 & 1.0 & 3.3 & 0.4 & 5.8 \\ 
   &  &  & 20\% & 10.7 & 10.7 & 7.8 & 5.8 & 0.8 & 3.6 & 0.3 & 5.9 \\ 
   \hline
Clayton & 0.0 & 0.05 & 10\% & 10.6 & 10.6 & 6.4 & 5.2 & 1.4 & 5.6 & 0.5 & 4.0 \\ 
   &  &  & 20\% & 11.2 & 11.2 & 7.4 & 7.0 & 1.5 & 7.2 & 0.5 & 4.3 \\ 
   &  & 0.20 & 10\% & 10.2 & 10.2 & 6.3 & 5.0 & 2.0 & 6.3 & 0.7 & 3.3 \\ 
   &  &  & 20\% & 11.4 & 11.4 & 7.4 & 7.6 & 2.4 & 7.1 & 0.8 & 3.5 \\ 
   &  & 0.40 & 10\% & 11.0 & 11.0 & 6.2 & 5.2 & 2.9 & 7.2 & 0.9 & 2.8 \\ 
   &  &  & 20\% & 14.2 & 14.2 & 8.5 & 8.3 & 3.6 & 9.1 & 1.3 & 3.3 \\ 
   \hline
 & 0.4 & 0.05 & 10\% & 10.4 & 10.4 & 6.8 & 5.5 & 0.9 & 6.0 & 0.5 & 5.9 \\ 
   &  &  & 20\% & 10.9 & 10.9 & 7.4 & 6.5 & 1.0 & 6.6 & 0.4 & 6.6 \\ 
   &  & 0.20 & 10\% & 10.3 & 10.3 & 6.3 & 4.8 & 1.4 & 5.3 & 0.6 & 4.5 \\ 
   &  &  & 20\% & 11.2 & 11.2 & 7.7 & 6.7 & 1.4 & 6.3 & 0.6 & 5.9 \\ 
   &  & 0.40 & 10\% & 11.1 & 11.1 & 6.4 & 5.1 & 2.1 & 5.7 & 0.8 & 4.1 \\ 
   &  &  & 20\% & 12.7 & 12.7 & 8.0 & 7.0 & 1.9 & 6.0 & 0.6 & 4.8 \\ 
   \hline
 & 0.8 & 0.05 & 10\% & 10.5 & 10.5 & 6.3 & 5.2 & 0.5 & 4.8 & 0.5 & 7.0 \\ 
   &  &  & 20\% & 12.1 & 12.1 & 8.5 & 7.4 & 0.5 & 6.0 & 0.4 & 8.3 \\ 
   &  & 0.20 & 10\% & 11.1 & 11.1 & 6.6 & 5.0 & 0.7 & 3.8 & 0.4 & 6.4 \\ 
   &  &  & 20\% & 10.7 & 10.7 & 7.9 & 6.3 & 0.5 & 4.3 & 0.4 & 6.8 \\ 
   &  & 0.40 & 10\% & 10.0 & 10.0 & 6.0 & 4.4 & 1.0 & 3.2 & 0.4 & 5.8 \\ 
   &  &  & 20\% & 10.4 & 10.4 & 7.1 & 5.7 & 0.9 & 3.5 & 0.4 & 6.1 \\ 
   \hline
\end{tabular}}
\end{table}

\begin{table}[!p]
\centering
\caption{Correlation between ALSPAC clinical phenotypes}
\label{tableau ALSPAC.1}
\begin{tabular}{|ccc|}
\hline
\multicolumn{2}{|c}{Phenotypes} & $\rho$ \\ \hline
  LDL & ApoB & 0.873 \\ 
  HDL & ApoA1 & 0.828 \\ 
  HDL & Trigl & -0.408 \\ 
  ApoB & Trigl & 0.188 \\ 
  HDL & ApoB & -0.183 \\ 
  ApoA1 & Trigl & -0.127 \\ 
  ApoA1 & ApoB & -0.073 \\ 
  LDL & Trigl & -0.036 \\ 
  HDL & LDL & -0.031 \\ 
  LDL & ApoA1 & 0.021 \\ 
   \hline
\end{tabular}
\end{table}

\clearpage
\section*{Appendix F. Supplementary Figures}

\begin{figure}[!h]
\centering
\includegraphics[width=1\textwidth]{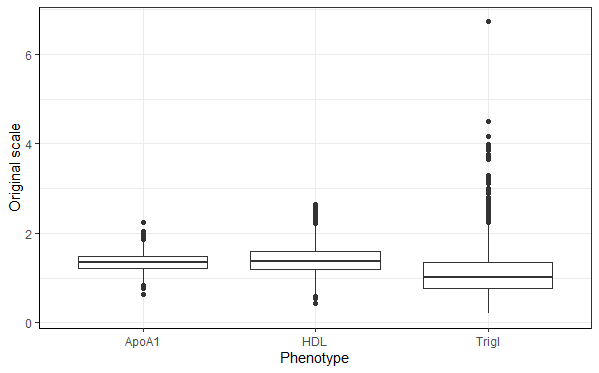}
\caption{Box plots for HDL, Trigl and ApoA1.}
\label{fig:ALSPAC.3}
\end{figure}

\begin{figure}[!h]
\centering
\includegraphics[width=1\textwidth]{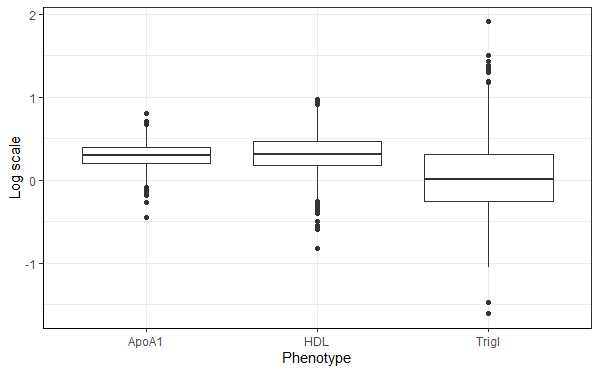}
\caption{Box plots for HDL, Trigl and ApoA1 after log-transformation.}
\label{fig:ALSPAC.4}
\end{figure}

\clearpage
\bibliographystyle{biorefs}
\bibliography{biblio}